\documentclass[apj,numberedappendix,revtex4,twocolappendix]{emulateapj}
\usepackage{natbib}
\usepackage{graphicx}
\usepackage[toc,page]{appendix}
\usepackage{natbib}
\usepackage{color}
\usepackage{threeparttable}
\usepackage{hyperref}
\usepackage{breakurl}
\begin{document}

\slugcomment{Draft version: \today}
\title{Constraining the Low-Mass Slope of the Star Formation Sequence at $0.5<z<2.5$}
\email{kate.whitaker@nasa.gov}
\author{Katherine E. Whitaker\altaffilmark{1,6}, Marijn Franx\altaffilmark{2}, Joel Leja\altaffilmark{3}, Pieter G. van Dokkum\altaffilmark{3}, 
Alaina Henry\altaffilmark{1,6}, \\ Rosalind E. Skelton\altaffilmark{4}, Mattia Fumagalli\altaffilmark{2}, 
Ivelina G. Momcheva\altaffilmark{3}, Gabriel B. Brammer\altaffilmark{5}, Ivo Labb\'{e}\altaffilmark{2},\\ 
Erica J. Nelson\altaffilmark{3}, Jane R. Rigby\altaffilmark{1}}
\altaffiltext{1}{Astrophysics Science Division, Goddard Space Flight Center, Code 665, Greenbelt, MD 20771, USA}
\altaffiltext{2}{Leiden Observatory, Leiden University, P.O. Box 9513, 2300 RA Leiden, The Netherlands}
\altaffiltext{3}{Department of Astronomy, Yale University, New Haven, CT 06520, USA}
\altaffiltext{4}{South African Astronomical Observatory, PO Box 9, Observatory, Cape Town, 7935, South Africa}
\altaffiltext{5}{ Space Telescope Science Institute, 3700 San Martin Drive, Baltimore, MD 21218, USA}
\altaffiltext{6}{NASA Postdoctoral Program Fellow}
\shortauthors{Whitaker et al.}
\shorttitle{Constraining the Low-Mass Slope of the Star Formation Sequence}

\begin{abstract}
We constrain the slope of the star formation rate ($\log\Psi$) to stellar mass ($\log\mathrm{M_{\star}}$) 
relation down to $\log(\mathrm{M_{\star}/M_{\odot}})=8.4$ ($\log(\mathrm{M_{\star}/M_{\odot}})=9.2$) 
at $z=0.5$ ($z=2.5$) with a mass-complete sample of 39,106 star-forming galaxies selected from the 3D-HST photometric
catalogs, using deep photometry in the CANDELS fields.
For the first time, we find that the slope is dependent on stellar mass, such that it is steeper at low masses 
($\log\mathrm{\Psi}\propto\log\mathrm{M_{\star}}$) than at high masses ($\log\mathrm{\Psi}\propto(0.3-0.6)\log\mathrm{M_{\star}}$).
These steeper low mass slopes are found for three different star formation indicators:
the combination of the ultraviolet (UV) and infrared (IR), calibrated from a stacking analysis
of Spitzer/MIPS 24$\mu$m imaging; $\beta$-corrected UV SFRs; and H$\alpha$ SFRs.
The normalization of the sequence evolves differently in distinct mass regimes as well: for galaxies
less massive than $\log(\mathrm{M_{\star}/M_{\odot}})<10$ the specific SFR ($\Psi/\mathrm{M_{\star}}$) is observed to be roughly self-similar
with $\Psi/\mathrm{M_{\star}}\propto(1+z)^{1.9}$, whereas more massive galaxies show a 
stronger evolution with $\Psi/\mathrm{M_{\star}}\propto(1+z)^{2.2-3.5}$ for $\log(\mathrm{M_{\star}/M_{\odot}})=10.2-11.2$.
The fact that we find a steep slope of the star formation sequence for the lower mass galaxies will help reconcile theoretical
galaxy formation models with the observations.  
\end{abstract}


\section{Introduction}
\label{sec:intro}

Galaxy surveys spanning the last 12 billion years of cosmic time have revealed a picture where 
the majority of star-forming galaxies follow a relatively tight relation between star formation rate ($\mathrm{SFR}\equiv\Psi$) and 
stellar mass (M$_{\star}$) \citep{Brinchmann04, Noeske07a, Elbaz07, Daddi07, Magdis10, Gonzalez10, Whitaker12b, Huang12}. 
The SFR increases with M$_{\star}$ as a power law ($\mathrm{\Psi}\propto\mathrm{M}_{\star}^{\alpha}$),
and these star-forming galaxies exhibit an intrinsic scatter of $\sim0.2-0.3$ dex that
is constant down to the completeness limits \citep{Whitaker12b, Speagle14}.
The observed relation suggests that prior to the shutdown of star formation, galaxy star formation histories
are predominantly regular and smoothly declining on mass-dependent timescales, rather than
driven by stochastic events like major mergers and starbursts.

The existence and tightness of the sequence argues for relatively steady star formation histories,
where the average specific star formation rates ($\mathrm{sSFR}\equiv\mathrm{\Psi/M_{\star}}$) of galaxies are observed to increase with redshift
as $\Psi/\mathrm{M_{\star}}\propto(1+z)^{3.4}$ \citep[e.g.][]{Oliver10}, 
with a flattening at $z>2$ \citep[e.g.,][]{Gonzalez10}.
Measurements of the slope $\alpha$ vary widely in the literature, ranging
between 0.2--1.2 \citep[see summary in ][]{Speagle14}.  However, after accounting for selection effects, the choice of stellar
initial mass function and the luminosity-to-SFR conversion, \citet{Speagle14} find a general consensus 
among star formation sequence observations across most SFR indicators, redshift ranges and stellar masses probed from
25 different studies,
reporting a $\sim0.1$ dex interpublication scatter.  It is now well established 
that there is a strong evolution in the normalization of the star formation sequence with redshift, and there
appears to be a consensus amongst slope measurements.  However, these studies are flux limited and thus target more 
massive galaxies.  Little is known about the evolution of the SFRs for low
mass galaxies beyond the low redshift Universe.  

Using a combination of UV and mid-infrared (mid-IR) calibrated 
SFRs (henceforth, UV+IR), \citet{Whitaker12b} suggested that there may 
be a curvature of the star-formation sequence, with a steeper slope at the low-mass end.  However, in that study
it is unclear how the effects
of sample incompleteness below 10$^{10}$ M$_{\odot}$ change the measured slope.  
Using a radio-stacking measurement of the SFR, \citet{Karim11} see similar trends,
but again are hampered by the mass completeness limits of the survey.

Probing the properties of the star formation sequence across a large range in stellar masses of galaxies
has only recently become feasible with the deep near infrared (NIR) high resolution
imaging from the Hubble Space Telescope (HST)
Wide Field Camera 3 (WFC3).  In particular, we exploit the deep WFC3 imaging provided by the
Cosmic Assembly Near-IR Deep Extragalactic Legacy Survey
\citep[CANDELS; ][]{Grogin11,Koekemoer11}, together with
a suite of ground and space-based observations across the entire electromagnetic
spectrum \citep[see ][for a summary of the optical and NIR photometry]{Skelton14}.

Although we now have a complete sample of galaxies at low stellar masses and out to high redshifts through the deep photometry 
available in the CANDELS fields, 
systematic uncertainties
remain when combining different SFR indicators.  
There does not exist a single SFR indicator that can probe the full dynamic range of stellar masses for individual galaxies \citep{Wuyts11a}.
Even with independent SFR indicators such as H$\alpha$ SFRs or rest-frame UV SFRs, deep IR observations are invaluable.
On average, only half of the light emitted from hot stars is in the UV relative to that absorbed by dust and re-emitted in the IR
for a galaxy with M$_{\star}$=10$^{9}$ M$_{\odot}$, decreasing to $<2\%$ at M$_{\star}$=10$^{11}$ M$_{\odot}$ \citep{Reddy06, Whitaker12b}.
\citet{Wuyts11a} show that dust correction methods, such as correcting for dust attenuation with $A_{V}$ measured from spectral energy distribution 
(SED) modeling or using UV continuum measurements, fail to recover the total amount of star 
formation in galaxies with high levels of star formation (e.g., $\Psi>100$ M$_{\odot}$ yr$^{-1}$).
However, the characteristic mass where the slope of the star formation sequence potentially changes
coincides with the limits of the deepest IR observations available.

The deepest IR data available across the legacy extragalactic fields \citep[e.g., Spitzer/MIPS 24$\mu$m imaging in the GOODS fields,][]{Dickinson03} 
only probes the average SFR down to $\sim10^{10}$ M$_{\star}$ at $z>1$, whereas the mass-completeness limits of the 3D-HST survey now 
extends about 1 decade lower \citep[e.g.,][]{Tal14, Skelton14}. Therefore, the best approach with the currently available data is 
through stacking analyses.  

In this paper, we demonstrate that we can measure the $\log\Psi-\log\mathrm{M_{\star}}$ relation down to 
the mass-completeness limits enabled by the deep imaging in the CANDELS fields without relying on calibrating different SFR indicators\footnote{In 
the \citet{Wuyts11a} SFR ladder, they calibrate the Spitzer 24$\mu$m and Hershel/PACS UV+IR SFRs at higher masses with SFRs from spectral energy
distribution modeling at lower masses to probe a broad range of stellar mass.}.
In the 3D-HST catalogs, we combine the deep 0.3-8$\mu$m photometry of the CANDELS fields with the HST/WFC3 G141 grism redshifts from the 3D-HST treasury program \citep{Brammer12}
and Spitzer/MIPS 24$\mu$m imaging.  

The paper is organized as follows: In Section~\ref{sec:data}, we present the details of the photometric catalogs, redshifts, stellar
masses and SFRs derived by the 3D-HST collaboration.  The observed star formation sequence is presented in Section~\ref{sec:sfr}, with direct measurements quantified
in Section~\ref{sec:quantify}.  We explore systematic uncertainties in the UV+IR SFR calibrations
in Section~\ref{sec:sysunc}, 
and compare with other SFR indicators in Section~\ref{sec:othersfrs}, such as H$\alpha$ SFRs and the UV SFR corrected for 
the shape of the rest-frame UV continuum.
In Section~\ref{sec:ssfr}, we present the mass-dependent evolution of the normalization to the observed star formation sequence.
Finally, we place these results
in the broader context of galaxy formation theories in Section~\ref{sec:broad}.  

In this paper, we use a \citet{Chabrier} initial mass function (IMF) and 
assume a $\Lambda$CDM cosmology with $\Omega_{M}$=0.3, $\Omega_{\Lambda}$=0.7, 
and $H_{0}$=70 km s$^{-1}$ Mpc$^{-1}$.  All magnitudes are 
given in the AB system.

\section{Data}
\label{sec:data}
\subsection{0.3-8$\mu$m Photometry and Grism Spectroscopy}

We exploit the exquisite HST/WFC3 and ACS imaging and spectroscopy over five well-studied 
extragalactic fields through the CANDELS and 3D-HST programs.  The fields are comprised of the
AEGIS, COSMOS, GOODS-North, GOODS-South, and the UKIDSS UDS fields, with a total area of $\sim$900 arcmin$^{2}$. 
A particular advantage of these fields is the wealth of publicly available imaging datasets in 
addition to the HST data, which makes it possible to construct the SEDs
for galaxies over a large wavelength range.  The number of optical to near-infrared (NIR) photometric broadband
and medium-bandwidth filters 
included in the \citet{Skelton14} photometric catalogs for each field ranges from 18 in UDS up to 44 in COSMOS. 
The sample used in this paper is selected from combined
$J_{\mathrm{F125W}}$+$H_{\mathrm{F140W}}$+$H_{\mathrm{F160W}}$ detection images,
with photometric redshifts and rest-frame colors determined with the EAZY code \citep{Brammer08}.
\citet{Skelton14} describe in full detail the 0.3--8$\mu$m photometric catalogs and data products
used herein.  All photometric catalogs are available through the 3D-HST
website\footnote{\url{http://3dhst.research.yale.edu/Data.html}}.

Where available, we combine the photometry with the spatially resolved low-resolution HST/WFC3 G141
grism spectroscopy to derive improved redshifts and emission line diagnostics.  
The $5\sigma$ continuum depth is $H_{\mathrm{F140W}}\sim23$.  \citet{Brammer12} introduce 
the 248--orbit 3D-HST NIR spectroscopic survey, whereas I. Momcheva et al., (in preparation) will present the full details of the 
data reduction and redshift analysis.  
For the purposes of this analysis, we select the ``best'' redshift to be the 
spectroscopic redshift, grism redshift or the photometric redshift, in this ranked order depending on the availability.  4\% of 
the final sample has a spectroscopic redshift, 12\% a grism redshift
 and 84\% a photometric redshift. 
The grism redshifts are only measured down to $H_{F160W}=23$ mag for the 3D-HST v4.0 internal release, whereas the public
release of the 3D-HST grism spectroscopy will measure grism redshifts for all objects.
Among galaxies more massive than 10$^{10}$ M$_{\odot}$, 13\% have a spectroscopic redshift,
38\% a grism redshift and 49\% a photometric redshift.
To determine the grism redshifts, we first compute a purely photometric redshift from the photometry,
using the EAZY code \citep{Brammer08}.  We then fit the full two-dimensional grism spectrum
separately with a combination of the continuum template taken from the EAZY fit and an
emission-line-only template with fixed line ratios taken from the Sloan Digital Sky Survey (SDSS) composite
star-forming galaxy spectrum of \citet{Dobos12}.  The final grism redshift, {\tt z\_grism},
is determined on a finely-sampled redshift grid with the photometry-only redshift probability
distribution function used as a prior.  This method is more flexible than that
originally described by \citet{Brammer12}, but the redshift precision is similar with
$\sigma\sim 0.0035(1+z)$.  

\subsection{24$\mu$m Photometry}
\label{sec:mips}

We derive Spitzer/MIPS 24$\mu$m photometric catalogs using the same methodology as
that described by \citet{Skelton14}. 
The Spitzer/MIPS 24$\mu$m images and weight maps in the AEGIS field are provided by the Far-Infrared Deep Extragalactic 
Legacy (FIDEL) survey\footnote{\url{http://irsa.ipac.caltech.edu/data/SPITZER/FIDEL/}}~\citep{Dickinson07}, COSMOS from the S-COSMOS survey \citep{Sanders07}, 
GOODS-N and GOODS-S from Dickinson et al. (2003),
and UDS from the Spitzer UKIDSS Ultra Deep Survey\footnote{\url{http://irsa.ipac.caltech.edu/data/SPITZER/SpUDS/}} (SpUDS; PI: J. Dunlop).
Unlike in \citet{Skelton14}, we generate new combined WFC3 detection images to remove the effects of varying HST/WFC3 point spread 
functions (PSFs) between the filters.  
We match the HST/WFC3 images to the $H_{\mathrm{F160W}}$ PSF and create new PSF-matched combined  
$J_{\mathrm{F125W}}$+$H_{\mathrm{F140W}}$+$H_{\mathrm{F160W}}$ detection images.
Due to the large Spitzer/MIPS 24$\mu$m PSF of
$\sim6^{\prime\prime}$\footnote{\url{http://irsa.ipac.caltech.edu/data/SPITZER/docs/mips/mipsinstrumenthandbook/50/}}, we then rebin the
combined PSF-matched detection images and segmentation maps by a 
factor of three to a 0.18$^{\prime\prime}$ pixel scale. We register the Spitzer/MIPS 24$\mu$m images 
to the higher resolution detection images using the IRAF {\tt wregister} tool.  

We use the Multi-resolution Object PHotometry oN Galaxy Observations (MOPHONGO) code developed by I. Labb\'{e} to perform the 
photometry on the low resolution MIPS images \citep[see earlier work by ][]{Labbe06, Wuyts07, Marchesini09, Williams09, Whitaker11}.  
The code uses a high-resolution image as a prior to model
the contributions from neighboring blended sources in the lower resolution image.  An additional shift map 
captures any small astrometric differences between the high resolution reference image and the low resolution
photometry of interest by cross-correlating the positions of objects.  We select all objects with a signal-to-noise (S/N) ratio greater than 6
when generating the shift map.  The code includes a background-subtraction
correction, fit on scales that are a factor of three larger than the 24$^{\prime\prime}$ convolution kernel tile size and rejects any pixels that are greater than $2\sigma$ 
outliers.  As the PSF can vary across the image, the code uses a position-dependent convolution
kernel that maps the higher resolution PSF to the lower resolution PSF.  A series of Gaussian-weighted
Hermite polynomials are fit to the Fourier transform of $\sim$20 pseudo point-sources across the 24$\mu$m images (effectively point-sources
at the resolution of MIPS).  Unlike the
automated method used by \citet{Skelton14}, these point-sources need to be hand selected in the 24$\mu$m mid-IR images as the majority of
point-sources cleanly selected in the NIR imaging are extremely faint in the mid-IR.  Aperture photometry is performed in a
3$^{\prime\prime}$ aperture radius, with corrections that account for the flux that falls outside of the aperture due to the large PSF size
and the contaminating flux from neighboring sources as determined
from the model.  We adopt an aperture correction of 20\% to account for the flux that falls outside of the 12$^{\prime\prime}$ tile radius, 
as taken from the MIPS instrument handbook.

\subsection{Stellar Masses}

Stellar masses are derived by fitting the 0.3-8$\mu$m 3D-HST photometric catalogs 
with stellar population synthesis templates using FAST \citep{Kriek09a}.
We fix the redshift to the ``best'' redshift, rank ordered as spectroscopic, grism, or photometric, 
as described above, whereas the stellar masses presented in \citet{Skelton14} are based on 
the photometric redshifts alone (or the spectroscopic redshift where available).
The models input to FAST are a grid of \citet{BC03} models that assume a \citet{Chabrier}
IMF with solar metallicity and a range of ages (7.6–-10.1 Gyr), 
exponentially declining star formation histories ($7<\tau<10$ in log years) and 
dust extinction ($0<A_{V}<4$), as described in \citet{Skelton14}. 
The dust content is parameterized by the extinction in the V-band following 
the \citet{Calzetti00} extinction law.

There is an increasingly significant contribution to the broadband flux from emission lines
for galaxies with high equivalent widths,
which can result in systematic overestimates in the stellar masses.
This effect has recently been found to be particularly important
for bright, blue galaxies discovered at $3<z<7$ \citep{Stark13b}.
Whereas \citet{Atek11} find up to a factor of 2 overestimate in the stellar masses
(or 0.3 dex) for the highest equivalent width galaxies, Pacifici et al. (in prep)
show that the contamination to the broadband flux is $<0.1$ dex for typical galaxies with 
$\log(\mathrm{M_{\star}/M_{\odot}})>10$ and $z<2$.
The correction will be a function of stellar mass itself and redshift due to the time evolution of the
strong correlation between equivalent width and M$_{\star}$ \citep[e.g.,][]{Fumagalli12}.

Therefore, we correct the stellar masses by taking into account these emission line contributions to the photometry. The details
of the photometric corrections are described in Appendix A. As shown in Figure~\ref{fig:comparemass}, the
stellar masses before and after emission line corrections are in good agreement for $\log(\mathrm{M_{\star}/M_{\odot}})>10$ 
but are systematically different at lower masses.  We use the corrected masses in this paper; using the uncorrected 
masses instead leads to small changes in the fitted slopes but does not change any of our conclusions.  Therefore,
similar studies of the average galaxy population that do not correct the stellar masses for emission-line contamination will not be introducing
significant biases to their analyses.

\subsection{UV+IR Star Formation Rates}
\label{sec:uvirsfr}

The SFRs are determined by adding the rest-frame UV light from massive stars to that
re-radiated in the FIR \citep[e.g.,][]{Gordon00}. We adopt a luminosity-independent 
conversion from the observed Spitzer/MIPS 24$\mu$m flux density to the total IR luminosity,
$\mathrm{L_{IR}} \equiv \mathrm{L(8-1000\mu m)}$,
based on a single template that is the log average of \citet{DH02} templates 
with $1<\alpha<2.5$\footnote{$\alpha$ represents the relative contributions of the different local templates, 
where $1<\alpha<2.5$ is the range where the model reproduces well the empirical spectra and IR color trends.}, 
following \citet{Wuyts08}, \citet{Franx08}, and \citet{Muzzin10}.
\citet{Wuyts11a} demonstrate that this luminosity-independent conversion from 24$\mu$m to the bolometric IR 
luminosity yields estimates that are in good median agreement with measurements from Herschel/PACS photometry,
successfully recovering the total amount of star formation in galaxies.

\begin{figure*}[t]
\leavevmode
\centering
\includegraphics[width=0.95\linewidth]{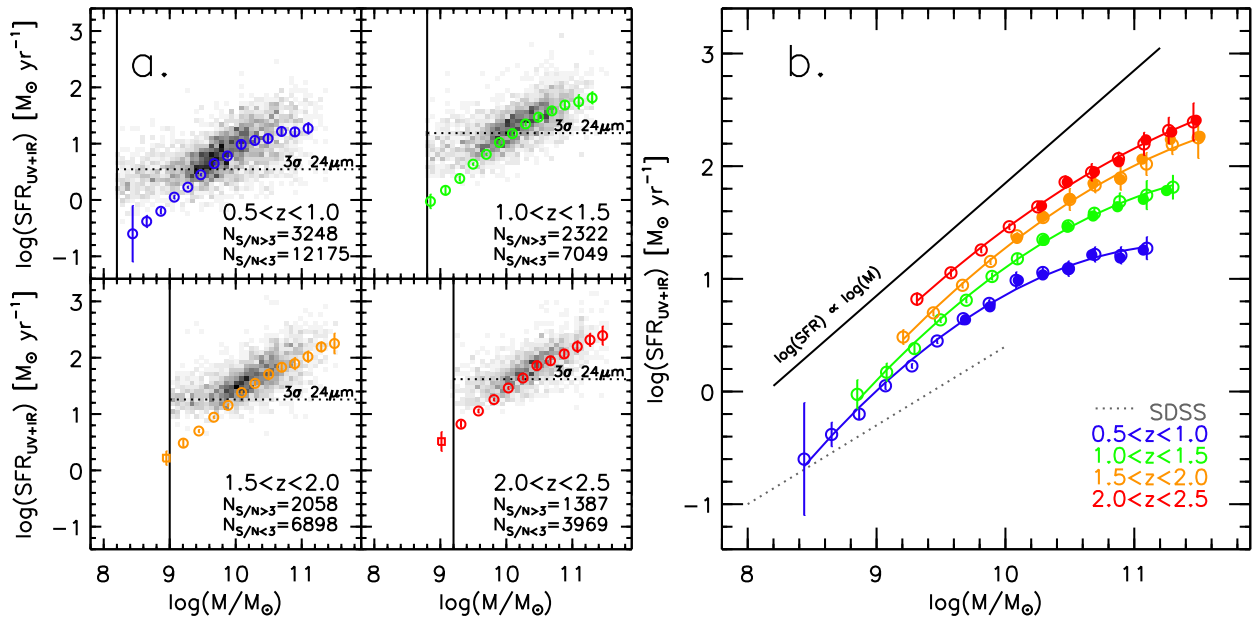}
\caption{The star formation rate as a function of stellar mass for star-forming galaxies.
Open circles indicate the UV+IR SFRs from a stacking analysis, 
with a second-order
polynomial fit above the mass completeness limits (solid vertical lines).  Open squares signify measurements
below the mass-completeness limits.
The running medians for individually detected objects in MIPS 24$\mu$m imaging with $\mathrm{S/N}>3$
(shown as a grey-scale density plot in the panel a, left) are indicated
with filled circles in the right panel and are color-coded by redshift.
The number of star-forming galaxies with $\mathrm{S/N}>3$ detections
in the 24$\mu$m imaging and those with $\mathrm{S/N}<3$ are indicated in the bottom right of each panel.  
The star formation sequence for star-forming galaxies is curved, with a constant slope of unity at $\log(\mathrm{M_{\star}/M_{\odot}})<10$ 
(solid black line in panel b is linear), whereas the slope at
the massive end flattens with $\alpha=0.3-0.6$ from $z=0.5$ to $z=2.5$.  
We show the SDSS curve (grey dotted line in panel b) from \citet{Brinchmann04} as it is one of the few measurements
which goes to very low mass, but it is based on another SFR indicator.}
\label{fig:sfr}
\end{figure*}

Building on the work of \citet{Bell05}, we make several simplifying assumptions when determining the rest-frame UV luminosities.
\citet{Bell05} estimated the $\mathrm{SFR_{UV}}$ using a calibration derived from the PEGASE stellar population models \citep{Fioc97}, 
assuming a 100 Myr old stellar population with constant SFR and a \citet{Kroupa01} IMF.  They estimate the total integrated 1216--3000$\mathrm{\AA}$
UV luminosity by using the 2800$\mathrm{\AA}$ rest-frame luminosity plus an additional factor of 1.5 to account for the UV spectral shape of a 
100 Myr old population with a constant SFR, where $\mathrm{L_{UV}} (1216-3000\mathrm{\AA}) = 1.5\nu\mathrm{L_{\nu, 2800}}$.
In our data, the rest-frame 2800$\mathrm{\AA}$ luminosity is determined from the best-fit template using the same methodology as
the rest-frame colors described in \citet{Brammer11}.  We choose to use the rest-frame 2800\AA\ luminosity instead of a
shorter wavelength as this ensures that the UV continuum will be sampled by at least 2 photometric bands for all galaxies.  

Assuming that $\mathrm{L_{IR}}$ reflects the bolometric luminosity of a completely obscured population of young 
stars and $\mathrm{L_{UV}}$ reflects the contribution of unobscured stars, \citet{Bell05} multiply $\mathrm{L_{UV}}$ 
by an additional factor of 2.2 to 
account for the unobscured starlight emitted shortward of 1216$\mathrm{\AA}$ and longward of 3000$\mathrm{\AA}$\footnote{We note that the SFR is always
dominated by the IR contribution, so in practice the assumptions for the derivation of $\mathrm{L_{IR}}$ will be most important.}.
Assuming a \citet{Chabrier} IMF, we therefore use the following luminosity to SFR ($\Psi$) conversion,

\begin{equation}
\mathrm{\Psi~[M_{\odot}~yr^{-1}] = 1.09\times10^{-10}(L_{IR}+2.2L_{UV})~[L_{\odot}]},
\label{eq:sfr}
\end{equation}

\noindent where $\mathrm{L_{IR}}$ is the bolometric IR (8--1000$\mu$m) luminosity
and $\mathrm{L_{UV}}$ is the total integrated rest-frame luminosity at 1216--3000$\mathrm{\AA}$ ($1.5\nu\mathrm{L_{\nu, 2800}}$).

\subsection{Sample Selection}
\label{sec:selection}

A standard method for discriminating star-forming galaxies from quiescent galaxies at high redshift is to select on the
rest-frame $U$--$V$ and $V$--$J$ colors \citep[e.g.,][]{Labbe05, Wuyts07, Williams09,Bundy10,Cardamone10b,Whitaker11,Brammer11,Patel12};
quiescent galaxies have strong Balmer/4000\AA\ breaks, characterized by red rest-frame $U$--$V$ colors and relatively blue rest-frame 
$V$--$J$ colors.  Following the two-color separations defined in \citet{Whitaker12a},
we select 58,973 star-forming galaxies at $0.5<z<2.5$ from the 3D-HST v4.0 catalogs\footnote{Essentially identical to
the publicly released catalogs available through \url{http://3dhst.research.yale.edu/Data.hmtl}, with the same catalog identifications
and photometry.}.  Of these, 39,106 star-forming galaxies are above the mass-completeness limits \citep{Tal14}.  Among the UVJ-selected star-forming 
galaxies with masses above the completeness limits, 22,253 have $\mathrm{S/N}>1$ MIPS 24$\mu$m detections (amongst which 9,015 have $\mathrm{S/N>3}$)
and 35,916 are undetected in MIPS 24$\mu$m photometry ($\mathrm{S/N}<1$)\footnote{Even though the SFR is dominated by the IR contribution, the
limiting factor here is the depth of the Spitzer/MIPS 24$\mu$m imaging.}.  The full sample of star-forming galaxies are considered in the
stacking analysis. Although we have not removed sources with X-ray detections in the following analysis, we estimate the 
contribution of active galactic nuclei (AGN) to the median 24$\mu$m flux densities in Section~\ref{sec:powerlaw}.

\begin{figure*}[t!]
\leavevmode
\centering
\includegraphics[width=0.95\linewidth]{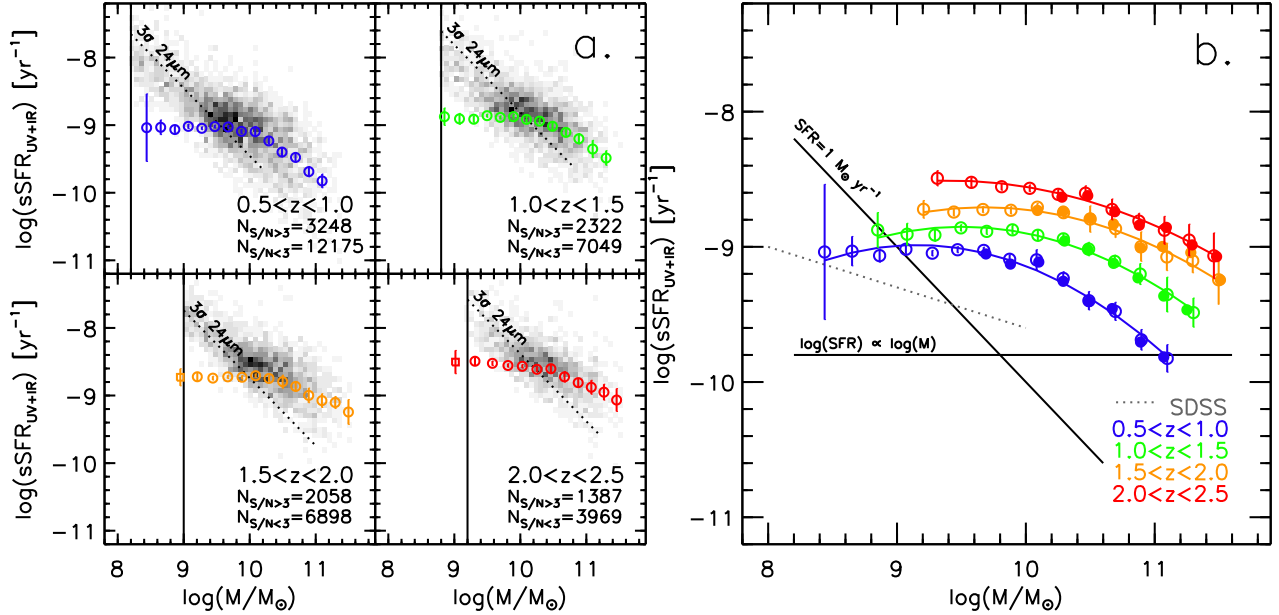}
\caption{Same data as presented in Figure~\ref{fig:sfr}, now showing the sSFR ($\equiv\mathrm{\Psi/M_{\star}}$)
as a function of stellar mass.  The sSFR-mass relation is flat for $\log(\mathrm{M_{\star}/M_{\odot}})<10$, 
with a tilt for the most massive galaxies indicating a stronger redshift evolution. }
\label{fig:ssfr}
\end{figure*}

\section{The Star Formation Sequence}
\label{sec:sfr}

Figure~\ref{fig:sfr} shows the star formation sequence, $\log\Psi$ as a function of $\log\mathrm{M_{\star}}$, in four redshifts bins from $z=0.5$ to $z=2.5$.
We use a single SFR indicator, the UV+IR SFRs described in Section~\ref{sec:uvirsfr}, probing over two decades in stellar mass.
The grey scale represents the density of points for star-forming galaxies selected in Section~\ref{sec:selection}
with $\mathrm{S/N}>3$ MIPS 24$\mu$m detections, totaling 9,015 star-forming galaxies over the full redshift range. 
Mass completeness limits are indicated by vertical lines.  The GOODS-N and GOODS-S fields have deeper
MIPS imaging ($3\sigma$ limit of $\sim$10$\mu$Jy) and HST/WFC3 $J_{F125W}$ and $H_{F160W}$ imaging ($5\sigma \sim 26.9$ mag),
whereas the other three fields have shallower MIPS imaging ($3\sigma$ limits of $\sim$20$\mu$Jy)
and HST/WFC3 $J_{F125W}$ and $H_{F160W}$ imaging ($5\sigma \sim 26.3$ mag).
The mass completeness limits in Figure~\ref{fig:sfr} correspond to the 90\% completeness limits derived by \citet{Tal14}, calculated by 
comparing object detection in the CANDELS/deep with a re-combined subset of the exposures which reach the depth of the CANDELS/wide
fields.  Although the mass completeness in the deeper GOODS-N and GOODS-S fields will extend to lower stellar masses, we adopt
the more conservative limits for the shallower HST/WFC3 imaging.

First, we look at the measurements for individual galaxies.  
The running median of the individual UV+IR measurements of the SFR are indicated with solid circles when the data are complete both in 
stellar mass and SFR (above the shallower data $3\sigma$ MIPS 24$\mu$m detection limit)\footnote{In the case of the $1.0<z<1.5$ and $1.5<z<2.5$ bins, 
the filled circles representing individual measurements
are limited by the $3\sigma$ 24$\mu$m completeness limits (horizontal dotted line, $\sim$20$\mu$Jy), which therefore makes it appear as though the higher redshift
sample extends to lower completeness limits due to the strongly evolving normalization.}. We consider all MIPS photometry in the median for
the individual UV+IR SFRs measurements (filled circles), even those galaxies intrinsically faint in the IR.
Only 1\% of the star-forming galaxies above the 20$\mu$Jy limit in each redshift bin have 24$\mu$m photometry with $\mathrm{S/N}<1$.

To leverage the additional decade lower in stellar mass that the CANDELS HST/WFC3 imaging enables us to probe for mass-complete
samples, next we stack the cleaned 24$\mu$m images for the full sample in stellar mass bins of 0.2 dex.  
Both galaxies that are detected and undetected in the MIPS 24$\mu$m photometry are included in the stacks, and the five CANDELS fields are
combined together.  In Appendix B, we repeat the stacking analysis in the five individual fields to demonstrate the differences due to
cosmic variance.
We subtract the average background in each individual stack, as measured in an annulus of 20--25$^{\prime\prime}$ radius. 
The photometry is extracted within a circular aperture of 3.5$^{\prime\prime}$ radius to increase the S/N ratio, with an aperture correction to
total of 2.57 taken from the MIPS handbook, assuming the background is determined from 20$^{\prime\prime}$ and beyond.  This method allows
us to reach far below the limits placed by the individual MIPS 24$\mu$m image depths, with the trade off that we are not able to 
measure the intrinsic scatter in the average $\log\Psi-\log\mathrm{M_{\star}}$ relation in this study.  
We note that the deep WFC3 prior allows us to cleanly extract the photometry below the standard 24$\mu$m confusion limit of $\sim 8 \mu$Jy.  
Clustering at scales smaller than the 24$\mu$m PSF size does not affect the photometry or stacking analyses \citep{Fumagalli13}.

The median stacked 24$\mu$m images for each 0.2 dex
stellar mass bin are added to the median UV luminosity of these galaxies. 
Results from stacking are indicated with open circles in Figure~\ref{fig:sfr}. 
Above the 3$\sigma$ MIPS 24$\mu$m limits, where a direct comparison is possible, we find that 
the stacks are in agreement with the running median of the individual UV+IR SFR measurements.  The average difference between the two measurements
is $0.00\pm0.02$ magnitudes.
The error bars for $\mathrm{L_{IR}}$ are derived from 50 bootstrap iterations of the 24$\mu$m stacking analysis for each stellar
mass bin.  These errors are added in quadrature to the 1$\sigma$ scatter in the $\mathrm{L_{UV}}$ values.
The solid black line in panel b indicates a slope of unity, where $\log\Psi$ is proportional to $\log\mathrm{M_{\star}}$.  We see in Figure~\ref{fig:sfr} 
that the slope is unity for less massive galaxies, but becomes shallower at the high-mass end.

Figure~\ref{fig:ssfr} presents the same data as that in Figure~\ref{fig:sfr}, however now showing the y-axis as the sSFR instead 
of SFR.  The sSFR varies with both redshift and mass (as indicated with the horizontal solid black line in panel b, for reference),
with a mass-dependence that becomes largest at the highest stellar masses.  This is 
another incarnation of the changing slope seen in Figure~\ref{fig:sfr}.  Similar to previous studies
in the literature, we find that the sSFR increases towards higher redshifts.  We explore the redshift evolution
of the sSFR in greater detail in Section~\ref{sec:ssfr}.

\begin{table}[t]
\centering
\begin{threeparttable}
    \caption{Polynomial Coefficients}\label{tab:polyfit}
    \begin{tabular}{lccc}
      \hline \hline
      \noalign{\smallskip}
    redshift range  & a & b & c \\
      \noalign{\smallskip}
      \hline
      \noalign{\smallskip}
      $0.5<z<1.0$  & $-27.40\pm1.91$ & $5.02\pm0.39$ & $-0.22\pm0.02$ \\
      $1.0<z<1.5$  & $-26.03\pm1.69$ & $4.62\pm0.34$ & $-0.19\pm0.02$ \\
      $1.5<z<2.0$  & $-24.04\pm2.08$ & $4.17\pm0.40$ & $-0.16\pm0.02$ \\
      $2.0<z<2.5$  & $-19.99\pm1.87$ & $3.44\pm0.36$ & $-0.13\pm0.02$ \\
      \noalign{\smallskip}
      \hline
      \noalign{\smallskip}
    \end{tabular}
    \begin{tablenotes}
      \small
    \item \emph{Notes.} Polynomial coefficients parameterizing the evolution of the $\log\Psi-\log\mathrm{M_{\star}}$
      relation from the median stacking analysis presented in Figure~\ref{fig:sfr} (see Equation~\ref{eq:polyfit}).
      When plotting the confidence intervals of the polynomial fits,                                                    
        note that the standard uncertainties in the polynomial coefficients are sign dependent.
    \end{tablenotes}
  \end{threeparttable}
\end{table}

\section{Quantifying the Star Formation Sequence}
\label{sec:quantify}

\subsection{Polynomial Fits}

We first fit the stacked $\log\Psi-\log\mathrm{M_{\star}}$ relation with a second-order polynomial for each redshift bin, considering only those
bins above the mass-completeness limits in the fit:

\begin{equation}
\log\Psi = a+b\log\left(\frac{\mathrm{M_{\star}}}{\mathrm{M_{\odot}}}\right)+c\log\left(\frac{\mathrm{M_{\star}}}{\mathrm{M_{\odot}}}\right)^{2}.
\label{eq:polyfit}
\end{equation}

The best-fit coefficients are presented in Table~\ref{tab:polyfit}.  The same relations are also presented in 
Figure~\ref{fig:ssfr}.  
In the case that one would want to 
re-derive the star formation sequence relations under a different set of 
assumptions (e.g., IMF, $\mathrm{L_{24\mu m}}$ to SFR conversion, etc), we provide the stacking analysis 
measurements in Table~\ref{tab:sfr}, including the measured SFRs, $\mathrm{L_{IR}}$,
and $\mathrm{L_{UV}}$ for all redshift and stellar mass bins for both the median and average stacks.  In Appendix C, we additionally provide
the measurements for a stacking analysis of all galaxies (including both quiescent and star-forming).

The coefficient of the second-order term ($c$ in Table~\ref{tab:polyfit}) is different
from zero at all redshifts, with a significance of $>6\sigma$ in each of 
the four redshift intervals. This demonstrates that the star formation sequence is 
not well described by a single slope. In the following subsection we will fit separate 
slopes to low and high mass galaxies.

\begin{table*}[ht]
\centering
\begin{threeparttable}
    \caption{Star Formation Sequence Data}\label{tab:sfr}
    \begin{tabular}{lrrrrrrrr}
      \hline \hline
      \noalign{\smallskip}
      & \multicolumn{1}{r}{$\log\mathrm{M_{\star}}$} &  \multicolumn{1}{c}{$\log\widetilde{\Psi}$} & \multicolumn{1}{c}{$\log\mathrm{\widetilde{L}_{IR}}$} & \multicolumn{1}{c}{$\log\mathrm{\widetilde{L}_{UV}}$} &  \multicolumn{1}{c}{$\log\overline{\Psi}$} & \multicolumn{1}{c}{$\log\mathrm{\overline{L}_{IR}}$} & \multicolumn{1}{c}{$\log\mathrm{\overline{L}_{UV}}$} & \multicolumn{1}{c}{$\beta$} \\
      \noalign{\smallskip}
      \hline
      \noalign{\smallskip}
      $0.5<z<1.0$  &  8.4 &  $-0.60\pm 0.50$ & $  7.69\pm  0.63$ & $  9.01\pm  0.01$ &  .................. &  .................. &  $9.04\pm  0.01$ &  -1.76 \\ 
             &  8.7 &  $-0.38\pm 0.11$ & $  8.83\pm  0.09$ & $  9.15\pm  0.01$ &  $-0.37\pm 0.11$ & $  8.75\pm  0.11$ & $  9.18\pm  0.01$ &  -1.74 \\ 
             &  8.9 &  $-0.20\pm 0.05$ & $  9.09\pm  0.05$ & $  9.32\pm  0.01$ &  $-0.19\pm 0.05$ & $  9.05\pm  0.05$ & $  9.34\pm  0.01$ &  -1.73 \\ 
             &  9.1 &  $ 0.05\pm 0.03$ & $  9.59\pm  0.03$ & $  9.47\pm  0.01$ &  $ 0.09\pm 0.03$ & $  9.64\pm  0.02$ & $  9.49\pm  0.01$ &  -1.67 \\ 
             &  9.3 &  $ 0.23\pm 0.02$ & $  9.85\pm  0.02$ & $  9.59\pm  0.01$ &  $ 0.27\pm 0.02$ & $  9.92\pm  0.01$ & $  9.61\pm  0.01$ &  -1.49 \\ 
             &  9.5 &  $ 0.45\pm 0.02$ & $ 10.17\pm  0.01$ & $  9.70\pm  0.01$ &  $ 0.53\pm 0.02$ & $ 10.28\pm  0.01$ & $  9.73\pm  0.01$ &  -1.32 \\ 
             &  9.7 &  $ 0.65\pm 0.02$ & $ 10.45\pm  0.02$ & $  9.76\pm  0.01$ &  $ 0.75\pm 0.02$ & $ 10.57\pm  0.01$ & $  9.81\pm  0.01$ &  -1.08 \\ 
             &  9.9 &  $ 0.78\pm 0.03$ & $ 10.63\pm  0.02$ & $  9.75\pm  0.02$ &  $ 0.88\pm 0.03$ & $ 10.75\pm  0.01$ & $  9.81\pm  0.02$ &  -0.84 \\ 
             & 10.1 &  $ 0.99\pm 0.07$ & $ 10.89\pm  0.02$ & $  9.72\pm  0.07$ &  $ 1.11\pm 0.07$ & $ 11.01\pm  0.01$ & $  9.82\pm  0.05$ &  -0.55 \\ 
             & 10.3 &  $ 1.06\pm 0.05$ & $ 10.96\pm  0.02$ & $  9.75\pm  0.02$ &  $ 1.18\pm 0.05$ & $ 11.09\pm  0.01$ & $  9.83\pm  0.01$ &  -0.42 \\ 
             & 10.5 &  $ 1.09\pm 0.07$ & $ 11.01\pm  0.02$ & $  9.65\pm  0.03$ &  $ 1.21\pm 0.07$ & $ 11.13\pm  0.01$ & $  9.75\pm  0.03$ &  -0.27 \\ 
             & 10.7 &  $ 1.22\pm 0.07$ & $ 11.13\pm  0.02$ & $  9.86\pm  0.01$ &  $ 1.36\pm 0.07$ & $ 11.28\pm  0.01$ & $  9.91\pm  0.01$ &  -0.37 \\ 
             & 10.9 &  $ 1.21\pm 0.08$ & $ 11.13\pm  0.02$ & $  9.79\pm  0.02$ &  $ 1.34\pm 0.08$ & $ 11.27\pm  0.01$ & $  9.87\pm  0.02$ &  -0.27 \\ 
             & 11.1 &  $ 1.27\pm 0.10$ & $ 11.19\pm  0.03$ & $  9.92\pm  0.03$ &  $ 1.40\pm 0.10$ & $ 11.32\pm  0.02$ & $  9.97\pm  0.02$ &  -0.42 \\
      \noalign{\smallskip}
      \hline
      \noalign{\smallskip}
      $1.0<z<1.5$  &  8.8 &  $-0.03\pm 0.13$ & $  9.20\pm  0.15$ & $  9.51\pm  0.01$ &  .................. &  .................. & $  9.55\pm  0.01$ &  -1.79 \\ 
             &  9.1 &  $ 0.17\pm 0.08$ & $  9.59\pm  0.08$ & $  9.64\pm  0.01$ &  $ 0.19\pm 0.08$ & $  9.58\pm  0.08$ & $  9.67\pm  0.01$ &  -1.76 \\ 
             &  9.3 &  $ 0.38\pm 0.05$ & $  9.96\pm  0.04$ & $  9.77\pm  0.01$ &  $ 0.37\pm 0.05$ & $  9.87\pm  0.05$ & $  9.80\pm  0.01$ &  -1.70 \\ 
             &  9.5 &  $ 0.64\pm 0.02$ & $ 10.36\pm  0.02$ & $  9.89\pm  0.01$ &  $ 0.68\pm 0.02$ & $ 10.41\pm  0.02$ & $  9.91\pm  0.01$ &  -1.55 \\ 
             &  9.7 &  $ 0.81\pm 0.02$ & $ 10.60\pm  0.02$ & $  9.95\pm  0.01$ &  $ 0.85\pm 0.02$ & $ 10.64\pm  0.01$ & $  9.98\pm  0.01$ &  -1.38 \\ 
             &  9.9 &  $ 1.02\pm 0.03$ & $ 10.88\pm  0.02$ & $  9.96\pm  0.03$ &  $ 1.12\pm 0.03$ & $ 10.99\pm  0.01$ & $ 10.02\pm  0.02$ &  -1.17 \\ 
             & 10.1 &  $ 1.18\pm 0.04$ & $ 11.06\pm  0.02$ & $ 10.02\pm  0.03$ &  $ 1.29\pm 0.04$ & $ 11.19\pm  0.01$ & $ 10.06\pm  0.03$ &  -0.97 \\ 
             & 10.3 &  $ 1.35\pm 0.04$ & $ 11.27\pm  0.02$ & $  9.90\pm  0.02$ &  $ 1.44\pm 0.04$ & $ 11.37\pm  0.01$ & $ 10.00\pm  0.02$ &  -0.71 \\ 
             & 10.5 &  $ 1.47\pm 0.05$ & $ 11.40\pm  0.02$ & $  9.92\pm  0.03$ &  $ 1.61\pm 0.05$ & $ 11.54\pm  0.01$ & $  9.99\pm  0.02$ &  -0.54 \\ 
             & 10.7 &  $ 1.58\pm 0.05$ & $ 11.52\pm  0.02$ & $  9.97\pm  0.01$ &  $ 1.70\pm 0.05$ & $ 11.64\pm  0.01$ & $ 10.01\pm  0.01$ &  -0.27 \\ 
             & 10.9 &  $ 1.69\pm 0.08$ & $ 11.63\pm  0.03$ & $ 10.01\pm  0.01$ &  $ 1.82\pm 0.08$ & $ 11.76\pm  0.02$ & $ 10.06\pm  0.01$ &  -0.20 \\ 
             & 11.1 &  $ 1.74\pm 0.13$ & $ 11.68\pm  0.05$ & $ 10.10\pm  0.03$ &  $ 1.80\pm 0.13$ & $ 11.74\pm  0.04$ & $ 10.16\pm  0.02$ &  -0.13 \\ 
             & 11.3 &  $ 1.81\pm 0.11$ & $ 11.76\pm  0.06$ & $ 10.00\pm  0.04$ &  $ 1.93\pm 0.11$ & $ 11.87\pm  0.05$ & $ 10.17\pm  0.03$ &  -0.21 \\ 
      \noalign{\smallskip}
      \hline
      \noalign{\smallskip}
      $1.5<z<2.0$  &  9.2 &  $ 0.48\pm 0.07$ & $ 10.00\pm  0.06$ & $  9.91\pm  0.01$ &  $ 0.47\pm 0.07$ & $  9.89\pm  0.08$ & $  9.94\pm  0.01$ &  -1.71 \\ 
             &  9.4 &  $ 0.70\pm 0.03$ & $ 10.35\pm  0.02$ & $ 10.02\pm  0.01$ &  $ 0.77\pm 0.03$ & $ 10.46\pm  0.02$ & $ 10.06\pm  0.01$ &  -1.59 \\ 
             &  9.7 &  $ 0.94\pm 0.02$ & $ 10.70\pm  0.02$ & $ 10.13\pm  0.01$ &  $ 0.97\pm 0.02$ & $ 10.73\pm  0.02$ & $ 10.16\pm  0.01$ &  -1.47 \\ 
             &  9.9 &  $ 1.15\pm 0.03$ & $ 11.00\pm  0.02$ & $ 10.15\pm  0.02$ &  $ 1.22\pm 0.03$ & $ 11.06\pm  0.02$ & $ 10.21\pm  0.01$ &  -1.25 \\ 
             & 10.1 &  $ 1.38\pm 0.03$ & $ 11.29\pm  0.02$ & $ 10.11\pm  0.02$ &  $ 1.45\pm 0.03$ & $ 11.35\pm  0.01$ & $ 10.20\pm  0.02$ &  -0.99 \\ 
             & 10.3 &  $ 1.54\pm 0.04$ & $ 11.47\pm  0.02$ & $ 10.08\pm  0.02$ &  $ 1.62\pm 0.04$ & $ 11.55\pm  0.02$ & $ 10.19\pm  0.02$ &  -0.78 \\ 
             & 10.5 &  $ 1.70\pm 0.10$ & $ 11.65\pm  0.02$ & $  9.99\pm  0.11$ &  $ 1.81\pm 0.10$ & $ 11.76\pm  0.02$ & $ 10.10\pm  0.09$ &  -0.61 \\ 
             & 10.7 &  $ 1.83\pm 0.07$ & $ 11.78\pm  0.02$ & $ 10.01\pm  0.03$ &  $ 1.91\pm 0.07$ & $ 11.85\pm  0.02$ & $ 10.09\pm  0.02$ &  -0.33 \\ 
             & 10.9 &  $ 1.90\pm 0.11$ & $ 11.84\pm  0.03$ & $ 10.01\pm  0.08$ &  $ 1.99\pm 0.11$ & $ 11.94\pm  0.02$ & $ 10.08\pm  0.07$ &  -0.24 \\ 
             & 11.1 &  $ 2.02\pm 0.10$ & $ 11.97\pm  0.04$ & $ 10.06\pm  0.03$ &  $ 2.13\pm 0.10$ & $ 12.08\pm  0.03$ & $ 10.09\pm  0.03$ &  -0.19 \\ 
             & 11.3 &  $ 2.19\pm 0.09$ & $ 12.14\pm  0.04$ & $ 10.23\pm  0.03$ &  $ 2.25\pm 0.09$ & $ 12.20\pm  0.04$ & $ 10.37\pm  0.02$ &  -0.14 \\ 
             & 11.5 &  $ 2.25\pm 0.18$ & $ 12.20\pm  0.10$ & $ 10.44\pm  0.04$ &  $ 2.30\pm 0.18$ & $ 12.25\pm  0.09$ & $ 10.40\pm  0.04$ &   0.06 \\ 
      \noalign{\smallskip}
      \hline
      \noalign{\smallskip}
      $2.0<z<2.5$  & 9.3 &  $ 0.82\pm 0.06$ & $ 10.44\pm  0.05$ & $ 10.18\pm  0.01$ &  $ 0.79\pm 0.06$ & $ 10.35\pm  0.06$ & $ 10.20\pm  0.01$ &  -1.58 \\ 
             &  9.6 &  $ 1.05\pm 0.03$ & $ 10.77\pm  0.03$ & $ 10.31\pm  0.01$ &  $ 1.05\pm 0.03$ & $ 10.76\pm  0.03$ & $ 10.33\pm  0.01$ &  -1.46 \\ 
             &  9.8 &  $ 1.26\pm 0.03$ & $ 11.04\pm  0.02$ & $ 10.40\pm  0.01$ &  $ 1.30\pm 0.03$ & $ 11.10\pm  0.02$ & $ 10.41\pm  0.01$ &  -1.29 \\ 
             & 10.0 &  $ 1.46\pm 0.03$ & $ 11.33\pm  0.01$ & $ 10.38\pm  0.02$ &  $ 1.51\pm 0.03$ & $ 11.38\pm  0.01$ & $ 10.44\pm  0.02$ &  -1.02 \\ 
             & 10.3 &  $ 1.64\pm 0.03$ & $ 11.55\pm  0.02$ & $ 10.30\pm  0.03$ &  $ 1.70\pm 0.03$ & $ 11.61\pm  0.01$ & $ 10.40\pm  0.02$ &  -0.84 \\ 
             & 10.5 &  $ 1.86\pm 0.05$ & $ 11.79\pm  0.02$ & $ 10.31\pm  0.04$ &  $ 1.95\pm 0.05$ & $ 11.89\pm  0.02$ & $ 10.34\pm  0.04$ &  -0.62 \\ 
             & 10.7 &  $ 1.95\pm 0.08$ & $ 11.89\pm  0.03$ & $ 10.17\pm  0.06$ &  $ 2.06\pm 0.08$ & $ 12.00\pm  0.02$ & $ 10.25\pm  0.05$ &  -0.36 \\ 
             & 10.9 &  $ 2.07\pm 0.06$ & $ 12.02\pm  0.03$ & $ 10.14\pm  0.02$ &  $ 2.13\pm 0.06$ & $ 12.08\pm  0.03$ & $ 10.23\pm  0.02$ &  -0.28 \\ 
             & 11.1 &  $ 2.20\pm 0.10$ & $ 12.15\pm  0.04$ & $ 10.12\pm  0.07$ &  $ 2.32\pm 0.10$ & $ 12.28\pm  0.03$ & $ 10.19\pm  0.06$ &  -0.07 \\ 
             & 11.3 &  $ 2.32\pm 0.12$ & $ 12.27\pm  0.06$ & $ 10.20\pm  0.05$ &  $ 2.42\pm 0.12$ & $ 12.38\pm  0.05$ & $ 10.23\pm  0.04$ &   0.16 \\ 
             & 11.5 &  $ 2.39\pm 0.17$ & $ 12.35\pm  0.10$ & $ 10.12\pm  0.11$ &  $ 2.60\pm 0.17$ & $ 12.56\pm  0.07$ & $ 10.24\pm  0.08$ &  -0.25 \\
      \noalign{\smallskip}
      \hline
      \noalign{\smallskip}
    \end{tabular}
    \begin{tablenotes}                                                                                                               
      \small                                                                                                                           
      \item \emph{Notes.} For more information and to download an ascii version of this table, 
        please visit \url{http://3dhst.research.yale.edu/Data.html}.
        Only star-forming galaxies selected based on their U-V and V-J rest-frame colors are considered here.
        Stellar masses are in units of M$_{\odot}$ and include a correction for emission-line contamination, as detailed in Appendix A.
        Star formation rates are in units of M$_{\odot}$ yr$^{-1}$.  
        Luminosities are in units of L$_{\odot}$. $\mathrm{\widetilde{L}_{IR}}$ ($\mathrm{\overline{L}_{IR}}$) is the bolometric 
        FIR luminosity as calibrated from the median (average) 24$\mu$m stacks, $\mathrm{\widetilde{L}_{UV}}$ ($\mathrm{\overline{L}_{UV}}$)
        is the median (average) rest-frame UV luminosity at 
        1216--3000$\mathrm{\AA}$ formally measured as $1.5\nu\mathrm{L_{\nu,2800}}$, and 
        $\beta$ is the average rest-frame UV continuum slope. Median stacks are signified as $\widetilde{\Psi}$
        (as presented in Figure~\ref{fig:sfr}),
        whereas the average stacks are noted as $\overline{\Psi}$.

    \end{tablenotes}                                                                                                                      
  \end{threeparttable}
\end{table*}

\subsection{Broken Power Law Fits}
\label{sec:powerlaw}

\begin{table}[t]
\centering
\begin{threeparttable}
    \caption{Broken Power Law Fits}\label{tab:lowfit}
    \begin{tabular}{lccc}
      \hline \hline
      redshift range ~~~~~ & $\mathrm{a_{low}}$ & $\mathrm{a_{high}}$ & b  \\
      \hline
      \noalign{\smallskip}
      $0.5<z<1.0$  & $0.94\pm0.03$ & $0.14\pm0.08$ & $1.11\pm0.03$ \\
      $1.0<z<1.5$  & $0.99\pm0.04$ & $0.51\pm0.07$ & $1.31\pm0.02$ \\
      $1.5<z<2.0$  & $1.04\pm0.05$ & $0.62\pm0.06$ & $1.49\pm0.02$ \\
      $2.0<z<2.5$  & $0.91\pm0.06$ & $0.67\pm0.06$ & $1.62\pm0.02$ \\
      \noalign{\smallskip}
      \hline
      \noalign{\smallskip}
    \end{tabular}
    \begin{tablenotes}
      \small
    \item \emph{Notes.} Broken power law coefficients parameterizing the evolution of the $\log\Psi-\log\mathrm{M_{\star}}$
      relation from the median stacking analysis for low and high-mass galaxies (Equation~\ref{eq:bpl}).  The characteristic mass is fixed at
      $\log(\mathrm{M_{\star}/M_{\odot}})=10.2$; $\mathrm{a_{low}}$ signifies the best-fit for galaxies below this limit and $\mathrm{a_{high}}$
      above this limit.
    \end{tablenotes}
  \end{threeparttable}
\end{table}

\begin{figure}[t]
\leavevmode
\centering
\includegraphics[width=\linewidth]{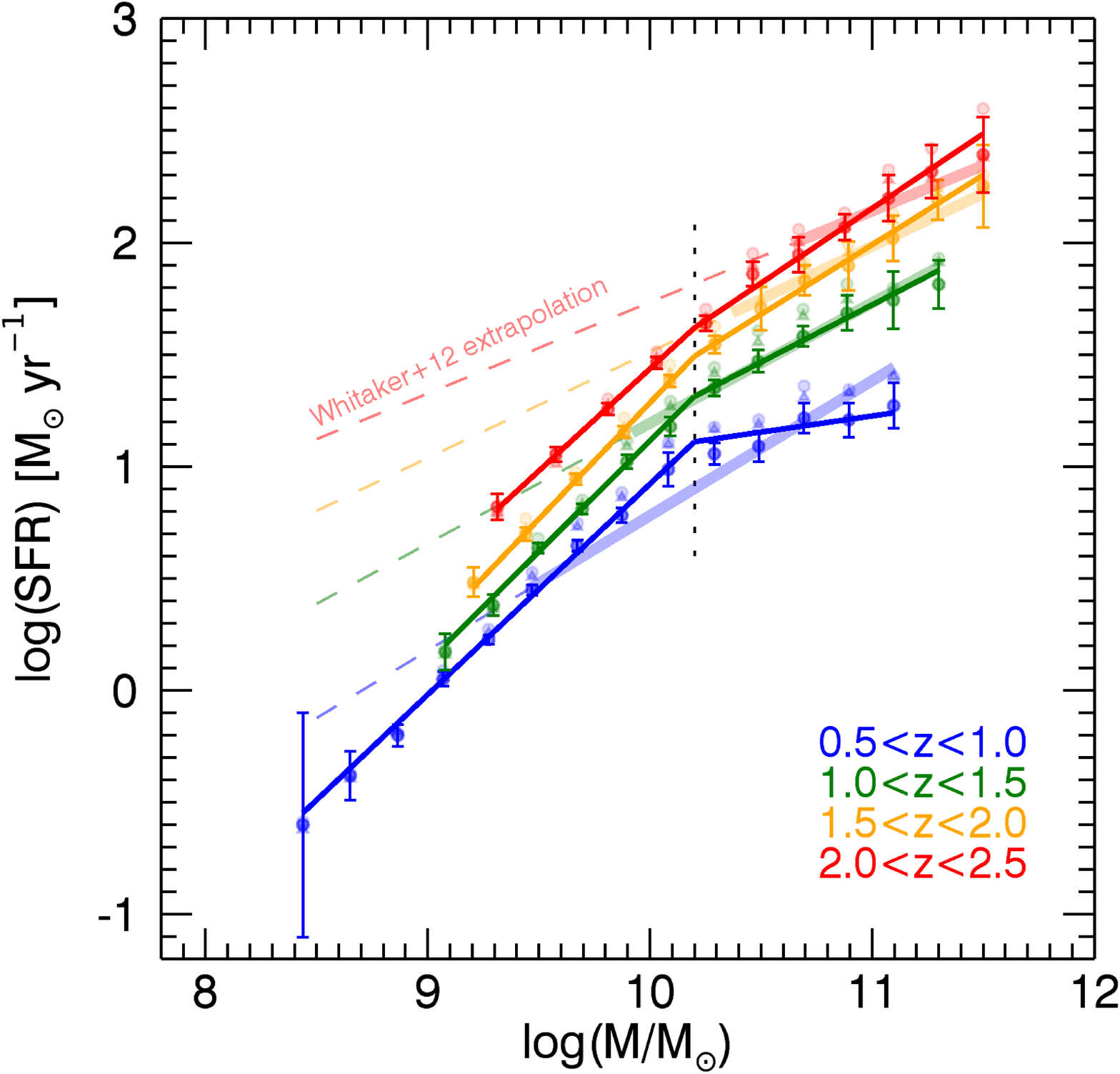}
\caption{The slope of the $\log\Psi-\log\mathrm{M_{\star}}$ relation is measured separately for star-forming galaxies more 
massive and less massive than $\log(\mathrm{M_{\star}/M_{\odot}})=10.2$, using the UV+IR SFRs derived through a stacking analysis.
The lighter color symbols indicate average stacks, whereas the darker color signify median stacks.  The triangles signify stacking analyses
where AGN candidates are removed.
We include the best-fit relations from \citet{Whitaker12b} 
for comparison, where the lighter color thick lines represent complete samples and the dashed lines indicate extrapolations 
into parameter space with no data. The \citet{Whitaker12b} relations are shifted down by 0.046 dex to account for the difference between the \citet{Chabrier} and \citet{Kroupa01} IMFs.}
\label{fig:slope1}
\end{figure}

\begin{figure}[t]
\leavevmode
\centering
\includegraphics[width=\linewidth]{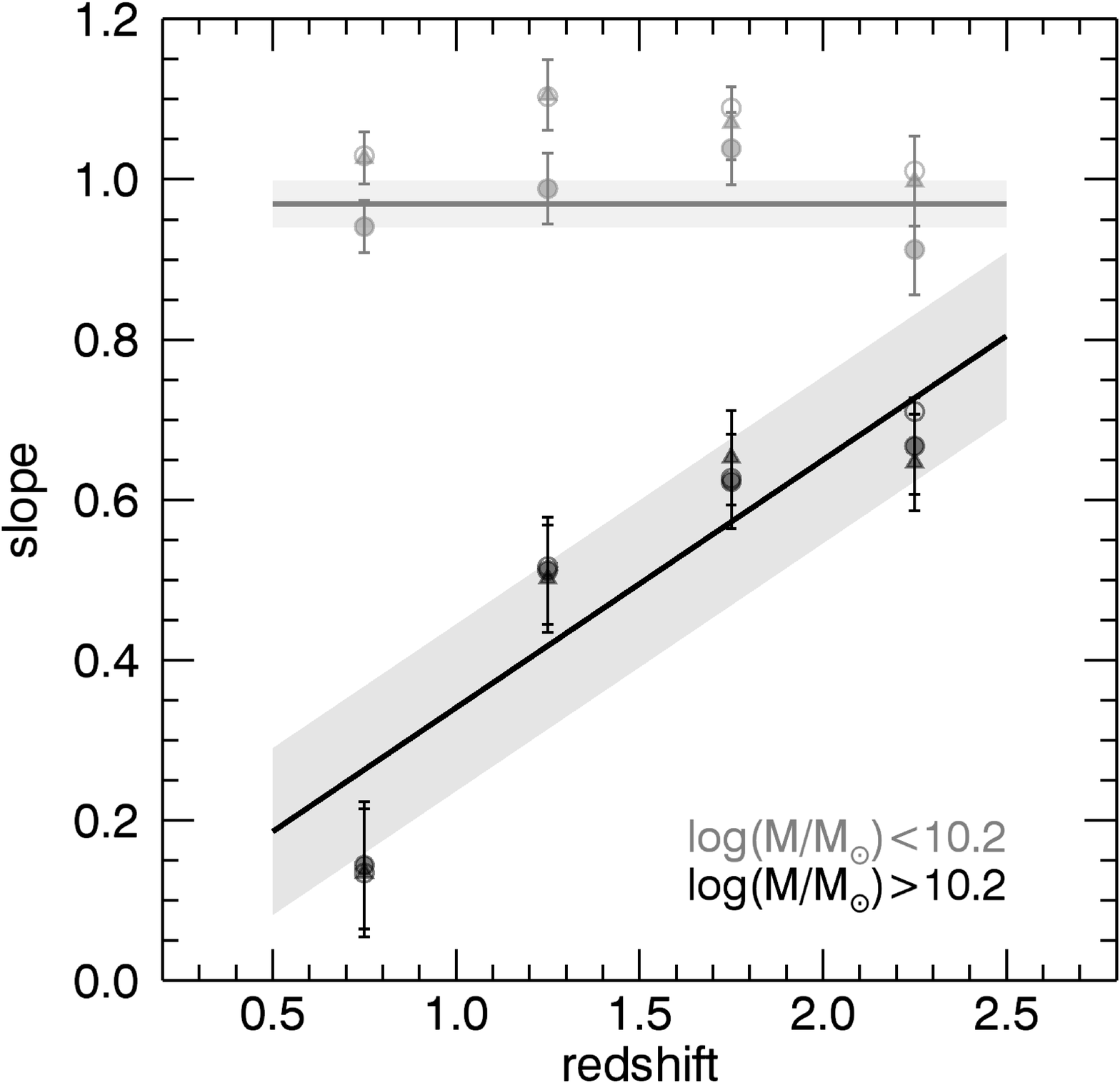}
\caption{Evolution of the slope of the $\log\Psi-\log\mathrm{M_{\star}}$ relation.  
The slope is roughly constant with redshift at $\sim1.0$ for low mass galaxies,
where as we see strong evolution in the slope for more massive galaxies, evolving from $\sim0.6$ at $z=2$ to $\sim0.3$ at $z=1$.
The open circles are measurements on the average stacks, whereas the filled circles represent median stacks.
The triangles signify stacking analyses where AGN candidates are removed.}
\label{fig:slope2}
\end{figure}

We fit the $\log\Psi-\log\mathrm{M_{\star}}$ relation with a broken power law, to 
independently quantify the behavior of the low mass galaxies and the
high mass galaxies.  Figure~\ref{fig:slope1} shows the best-fit relations
using the UV+IR SFRs derived in the stacking analysis with stellar masses corrected for emission line contamination.
The broken power law is parameterized as,

\begin{equation}
\log\Psi = a\left[\log\left(\frac{\mathrm{M_{\star}}}{\mathrm{M_{\odot}}}\right)-10.2\right]+b,
\label{eq:bpl}
\end{equation}

\noindent where the value of $a$ is different above and below the characteristic mass of $\log(\mathrm{M_{\star}/M_{\odot}})=10.2$.
If we instead allow the characteristic mass to vary, the best-fit value ranges from $\log(\mathrm{M_{\star}/M_{\odot}})=10.0\pm0.1$ at $0.5<z<1.0$,
to $\log(\mathrm{M_{\star}/M_{\odot}})=10.2\pm0.1$ at $1.0<z<2.0$, up to 
$\log(\mathrm{M_{\star}/M_{\odot}})=10.5\pm0.3$ at $2.0<z<2.5$.  Fixing the characteristic mass to $\log(\mathrm{M_{\star}/M_{\odot}})=10.2$ does not 
significantly effect the measured redshift evolution of the slope. 
The best-fit coefficients are presented in Table~\ref{tab:lowfit}.
The slope measurements at the high-mass end are consistent with the NMBS data analysis \citep{Whitaker12b}, shown in lighter colors for reference. 
However, we note that the slope evolution in Equation 2 of \citet{Whitaker12b}
does not formally agree due to the different stellar mass ranges considered in their linear fits (thick solid lines in Figure~\ref{fig:slope1}).
The dashed lines in Figure~\ref{fig:slope1} indicate extrapolations that enter parameter space where no data are available for NMBS.

We show the evolution of the measured slope for low mass ($\log(\mathrm{M_{\star}/M_{\odot}})<10.2$) and high mass ($\log(\mathrm{M_{\star}/M_{\odot}})>10.2$)
galaxies in Figure~\ref{fig:slope2}.  The data are consistent with no evolution in the slope for low mass galaxies,
with an average measured value of $\alpha=0.97\pm0.06$ (grey solid line)
\footnote{The formal best-fit for the redshift evolution 
of the low mass slope is $\alpha(z) = 0.95\pm0.05 + (0.02\pm0.04)z$.}

On the other hand, we observe a strong
evolution in the slope at the high mass end of the $\log\Psi-\log\mathrm{M_{\star}}$ relation.  We measure a slope of $\alpha\sim0.7$ at $z>2$, 
whereas the relation becomes shallower with time reaching a value of $\alpha\sim0.3$ by $z\sim1$.  
We fit the high-mass slopes as a function of redshift, finding 
a strong evolution in the slope for massive galaxies ($\log(\mathrm{M_{\star}/M_{\odot}})>10.2$), with the following best-fit relation:

\begin{equation}
\alpha(z) = 0.03\pm0.10 + (0.31\pm0.06)z.
\label{eq:highmass}
\end{equation}

To test if the changing slope of the star formation sequence could be partly a result of contamination due to AGN,     
we redo our stacking analyses after removing AGN candidates.  Spitzer/IRAC selection is a powerful tool for identifying luminous AGNs. As all five
CANDELS fields have uniform coverage with deep Spitzer/IRAC imaging, we remove all sources that fall within the revised IRAC color-color selections presented
in \citet{Donley12}.  The \citet{Donley12} selection is more restrictive than the standard ``wedge'' \citep[e.g., ][]{Stern05}, such that distant
star-forming galaxies are not removed unnecessarily.   The adopted AGN selection criteria is therefore optimized for the present dataset.
We find that removing AGN candidates from the sample does not significantly change the measured slope of the star formation sequence.  The
change in SFR varies smoothly from an overestimate of order 0.02--0.06 mag at the lowest masses when not removing AGN candidates, to an underestimate of
0.08--0.1 mag at the highest masses.  The results from the stacking analyses with AGN candidates removed are indicated by triangles in Figures~\ref{fig:slope1}
and \ref{fig:slope2}.
Although the average low-mass slope is higher with a measured value of $\alpha=1.06\pm0.06$, both the low-mass and high-mass slopes agree 
within the formal error bars.

In this section we have quantified the behavior of the slope of the star formation
sequence in different mass intervals.
We find that on average, emission line contamination 
to the stellar masses cannot account for the different slopes of low and high masses, as detailed in Appendix A.
Although we correct for emission line contamination, we have not accounted
for contamination to the broadband fluxes due to the nebular continuum here.  This effect is only significant where the
fraction of nebular gas relative to stars is high, or
$W_\mathrm{H\alpha}\gtrsim300$ \citep[e.g.,][]{Izotov11}.  There may therefore be additional contamination of our stellar masses
due to the nebular continuum for $\log(\mathrm{M_{\star}/M_{\odot}})<9.5$ at $z>2$.  In practice, the lowest mass bin above our mass-completeness 
limits at $2.0<z<2.5$ may be slightly overestimated.  
In the next section, we explore the effects of the assumed calibrations to convert the observed 24$\mu$m flux density and the 
rest-frame 2800$\mathrm{\AA}$ luminosities into SFRs.

\begin{figure}[t!]
\leavevmode
\centering
\includegraphics[width=\linewidth]{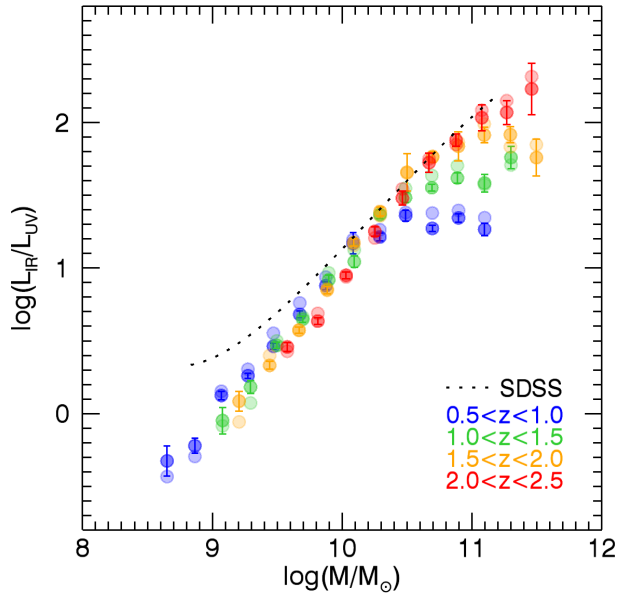}
\caption{The infrared excess ($\mathrm{IRX}\equiv\mathrm{L}_{\mathrm{IR}}$/L$_{\mathrm{UV}}$) 
as a function of stellar mass for the median and average stacks, darker and lighter color
symbols, respectively.  More massive star-forming galaxies have higher $\log\mathrm{(L_{IR}/L_{UV})}$ ratios, suggesting
increasing amounts of dust attenuation. There is surprisingly little evolution in the average $\log\mathrm{(L_{IR}/L_{UV})}$--$\log\mathrm{M_{\star}}$ relation
at $\log(\mathrm{M_{\star}/M_{\odot}})<10.5$.
There exists a trend at $\log(\mathrm{M_{\star}/M_{\odot}})>10.5$ for more dust per fixed stellar mass towards higher redshift,
but we note that the relation itself has a relatively large intrinsic scatter.  The dotted black line indicates the inferred
$\log\mathrm{(L_{IR}/L_{UV})}$ ratios from Balmer decrement measurements with SDSS from \citet{GarnBest10}.}
\label{fig:LIR_LUV}
\end{figure}

\section{Uncertainties in Star Formation Rates}
\label{sec:sysunc}

Here, we explore the uncertainties in the UV+IR star formation rates in greater detail.
In particular, we explore the implications of the dependence of the rest-frame IR to UV bolometric 
luminosity ratios (Section~\ref{sec:LIR_LUV}) and the gas-phase metallicity (Section~\ref{sec:metallicity}) on
stellar mass.

\subsection{Stellar Mass Dependence of L$_{\mathrm{IR}}$/L$_{\mathrm{UV}}$}
\label{sec:LIR_LUV}

We observe a curvature of the $\log\Psi-\log\mathrm{M_{\star}}$ relation towards lower sSFRs for the most massive
star-forming galaxies.  The implication is that the most massive star-forming galaxies have older stellar populations,
as the inverse of the sSFR defines a timescale for the formation of the stellar population of a galaxy.
However, the evolution measured for the mass dependency of the $\log\Psi-\log\mathrm{M_{\star}}$ relation 
in Figure~\ref{fig:sfr} relies on a proper accounting of the dust content of the galaxies.

The infrared excess ($\mathrm{IRX}\equiv\mathrm{L}_{\mathrm{IR}}$/L$_{\mathrm{UV}}$) probes the amount of dust in a galaxy, and has been
shown to be a strong function of stellar mass \citep[e.g.,][]{Reddy06, Reddy10, Whitaker12b}.
Figure~\ref{fig:LIR_LUV} shows the IRX ratio as derived from the stacking analysis for the different stellar mass bins,
including error bars from the bootstrap analysis. Although there is a large intrinsic scatter in this ratio \citep[e.g.,][]{Whitaker12b},
we see remarkably little redshift evolution in the average ratio below $\log(\mathrm{M_{\star}/M_{\odot}})=10.5$.

To our knowledge, there does not exist a direct measurement
of IRX as a function of stellar mass for local galaxies. 
The dotted black line in Figure~\ref{fig:LIR_LUV} is derived from the Balmer decrement measurements
from the SDSS sample presented in \citet{GarnBest10}.
We assume the \citet{Calzetti00} dust extinction law to convert the measured H$\alpha$ extinction to the extinction at 1600$\mathrm{\AA}$.
\citet{Meurer99} provide an empirical relation between A$_{1600}$ and IRX$_{1600}$.  We note that we are measuring IRX$_{\mathrm{UV}}$,
as calculated from $1.5\nu\mathrm{L}_{\nu2800}$, which should be roughly equivalent \citep{Kennicutt98}.
We find that the IRX ratio inferred from SDSS data is about 0.1 dex higher than the ratio we measure
in this work at $\log\mathrm{M_{\star}}=9.5-11$ M$_{\odot}$.  The offset of up to $\sim0.5$ dex at $\log\mathrm{M_{\star}}\lesssim9$ 
may result because the
\citet{GarnBest10} sample is selected on the basis of emission lines being detected, which could introduce
selection biases.
Although the overall offset of $\sim0.1$ dex may be due to calibration uncertainties, it is more likely related to the assumed
conversions to convert the direct SDSS Balmer decrement measurements to an inferred IRX ratio.
In our data, we find no evolution with redshift from $z=0.5$ to $z=2.5$ for $\log(\mathrm{M_{\star}/M_{\odot}})<10.5$.
There appears to be a trend for higher IRX ratios with increasing redshift for $\log(\mathrm{M_{\star}/M_{\odot}})>10.5$.  However, the
inferred SDSS IRX ratio is inconsistent with this evolution towards lower IRX ratios at higher masses.
If there are systematic errors in our calibration of $\mathrm{L_{IR}}$ to a SFR, then the effect of this error
will depend on mass.  In particular, if the conversion from 24$\mu$m to $\mathrm{L_{IR}}$ underestimated the luminosity, it would
introduce a trend similar to Figure~\ref{fig:sfr}.

\subsection{Effect of Gas-Phase Metallicity on 24$\mu$m Flux Density}
\label{sec:metallicity}

The gas-phase metallicities of galaxies are observed to be a strong function of their stellar mass \citep[e.g.,][]{Tremonti04,Erb06,Mannucci10,Zahid11,Steidel14}, where
less massive galaxies have fewer metals.  This relation has several implications with respect to the calibration
of SFR indicators over a broad range in stellar mass.
The Spitzer/MIPS 24$\mu$m imaging captures strong aromatic emission features, thought to be due to
polycyclic aromatic hydrocarbons (PAHs).  
PAH formation and destruction is sensitive to the metallicity of a galaxy;
observations show a metallicity-dependence of the PAH abundance in galaxies 
\citep[e.g.,][]{Engelbracht05, Madden06}, with a paucity of PAH detection in low metallicity galaxies 
\citep[$Z<0.1Z_{\odot}$; e.g.,][]{Hunt10}.  

Although SFRs derived from Spitzer/MIPS 24$\mu$m imaging are in excellent agreement with 
Herschel measurements \citep{Wuyts11a}, this indicator is not
well-calibrated in low mass galaxies.  Furthermore, the fraction of IR light from dust heated by old stars remains poorly
constrained, a problem that is likely most relevant in more massive galaxies \citep{Kennicutt98, Rieke09}.  
Taking the agreement with Herschel FIR measurements at face value, we only consider the former problem here;
the reliability of SFRs 
extrapolated from the empirical Spitzer/MIPS 24$\mu$m 
calibration is unknown for low mass galaxies.
\citet{Hunt10} show that although the fraction of PAH
emission normalized to the total IR luminosity is considerably smaller in metal-poor galaxies,
they show signs of a harder radiation field through a deficit 
of small PAHs.  The weaker PAH emission may be offset by the harder radiation field, conspiring to minimize potential
mass-dependent systematic effects when using the rest-frame 6--14$\mu$m to infer the IR SFR (where we probe $\sim7\mu$m at $z=2.5$ and 
just outside the PAH window at $\sim16\mu$m for $z=0.5$).  However, \citet{Engelbracht05} observe an abrupt shift in the mean 8-to-24$\mu$m
flux density ratio between 0.3 and $0.2Z_{\odot}$ due to a decrease in the 8$\mu$m flux density.

The paucity of PAH features is observed for metallicities less than 0.1--0.2$Z_{\odot}$. 
\citet{Zahid11} and \citet{Erb06} show that the
average metallicity is also $>0.1Z_{\odot}$ at least above $\log(\mathrm{M_{\star}/M_{\odot}})$ of 9.2 and 9.5 at $z=0.8$ and $z>2$, respectively.
Although the strength of the PAH features will depend on the metallicity, only our lowest mass bins could significantly suffer 
from 24$\mu$m luminosity to SFR calibration issues due to a complete lack of PAH features altogether. 
The steeper slope of the $\log\Psi-\log\mathrm{M_{\star}}$ relation is still observed at $\log(\mathrm{M_{\star}/M_{\odot}})=9.5-10$,
so we suspect that the measurements presented in Figure~\ref{fig:slope2} should therefore hold irrespective of unknown metallicity effects.
Future studies of gravitationally lensed low-mass systems may shed light on this matter.

As we cannot formally account for the systematic uncertainties in the $\mathrm{SFR_{IR}}$ introduced at low stellar masses due to changing
gas-phase metallicities, we instead
derive SFRs that are independent of the 24$\mu$m to bolometric IR luminosity conversion in the following section.  
In the absence of FIR data, we can 
estimate the total SFR from the measured UV luminosity by assuming a correlation between the slope of the UV continuum and dust extinction, and 
from H$\alpha$ emission.

\begin{figure}[t]
\leavevmode
\centering
\includegraphics[width=\linewidth]{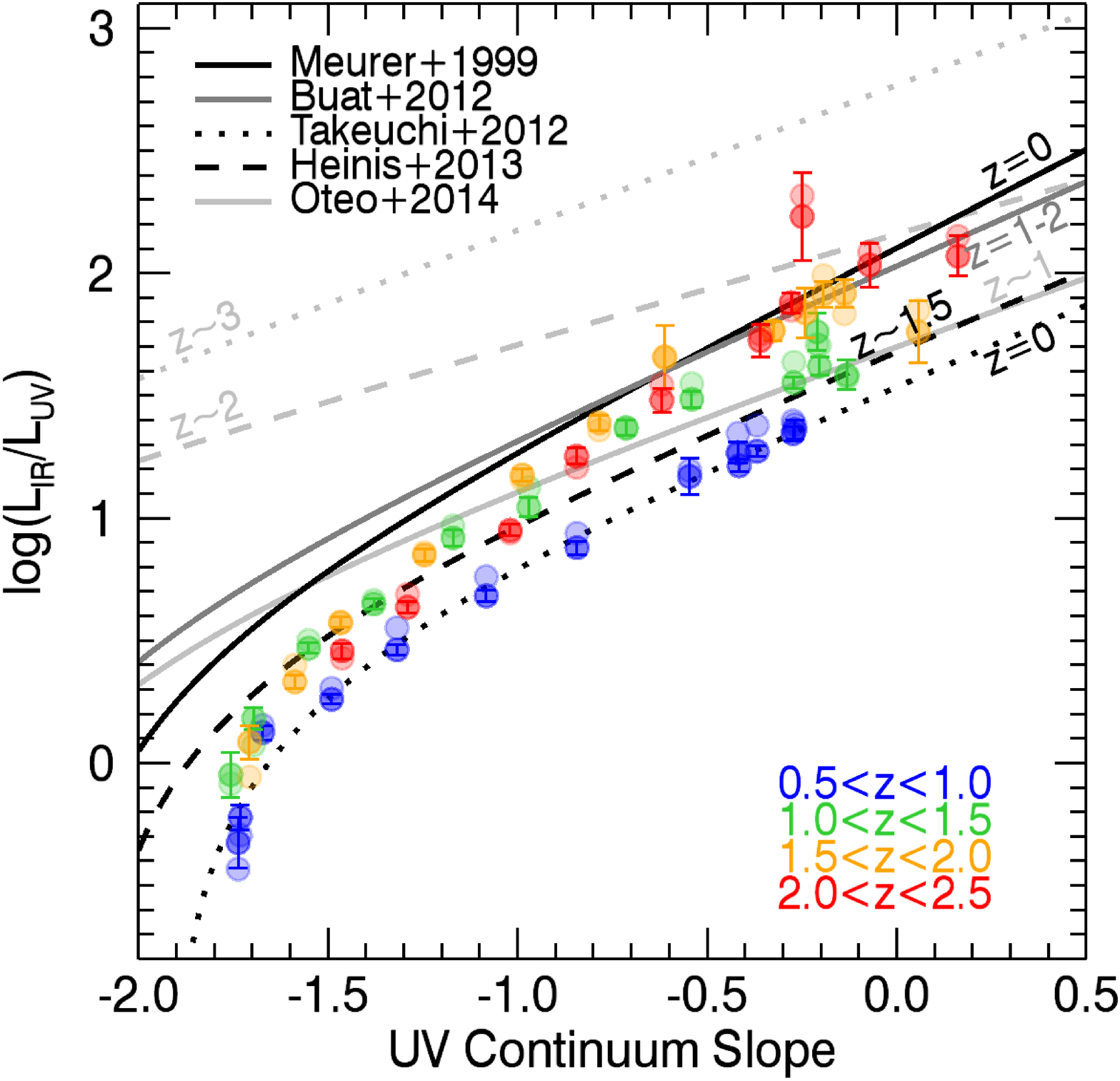}
\caption{The infrared excess ($\mathrm{IRX}\equiv\mathrm{L}_{\mathrm{IR}}$/L$_{\mathrm{UV}}$) as a function of the slope of the UV continuum,
where probe quantities probe the amount dust in a galaxy.  High IRX ratios and shallower UV continuum slopes are indicative of higher dust extinction.
The data is consistent with little evolution in the IRX-$\beta$ correlation at $1.0<z<2.5$, with an offset towards lower IRX ratios
for our lowest redshift bin of $0.5<z<1.0$. We compare our self-consistent measurements of this correlation to various parameterizations
presented in the literature.  The lighter color symbols represent the ratio from average stacks, whereas the darker color symbols 
were measured from median stacks.}
\label{fig:IRX_beta}
\end{figure}

\section{Other Star Formation Indicators}
\label{sec:othersfrs}

Adding the rest-frame UV light of massive stars to that re-radiated at FIR wavelengths is considered the best practice
to obtain reliable total SFRs \citep[e.g.,][]{Speagle14}, but we can also consider alternative SFR indicators that do not suffer from the same calibration 
uncertainties at low stellar masses.  
With the 3D-HST photometric catalogs, we can derive total SFRs from L$_{\mathrm{UV}}$ with a $\beta$ dust correction\footnote{The rest-frame UV spectral slope $\beta$
is determined from a power-law fit of the form $f_{\lambda}\propto\lambda^\beta$.}, and also
measure L$_{\mathrm{H}\alpha}$ (for a smaller redshift range where H$\alpha$ is covered by the WFC3/G141 grism). 

\subsection{Dust corrections to the UV Continuum}
\label{sec:UVcontinuum}

The rest-frame UV spectrum is nearly flat in L$_{\nu}$ over the wavelength range of 1200--3000\AA, which allows
us to express the conversion from the observed 2800\AA\ luminosity to the integrated UV luminosity
in a relatively simple form \citep{Kennicutt98}.  This conversion implicitly assumes that
galaxies continuously form stars over timescales of 100 Myr or longer.
As we are probing the average SFR of galaxy
populations as a whole, this assumption is probably justified.
To test our assumption of solar metallicity, we compare spectra produced from the stellar
synthesis code Starburst~99 \citep{Starburst99, Vazquez05, Leitherer10}. We find that
the correction to the rest-frame 2800\AA\ luminosity
to account for the slope between 1216--3000\AA\ increases by only 2\% when instead
assuming continuous star-formation over a 100 Myr with a \citet{Kroupa01} IMF and 0.4$Z_{\odot}$.
A higher metallicity of 2$Z_{\odot}$ results in a slightly shallower UV continuum slope and a decrease in the correction
to the rest-frame 2800\AA\ luminosity by 7\%.  Thus, our calculation of
the rest-frame UV luminosity does not strongly depend on our assumption of solar metallicity.

\begin{figure*}[t]
\leavevmode
\centering
\includegraphics[width=0.85\linewidth]{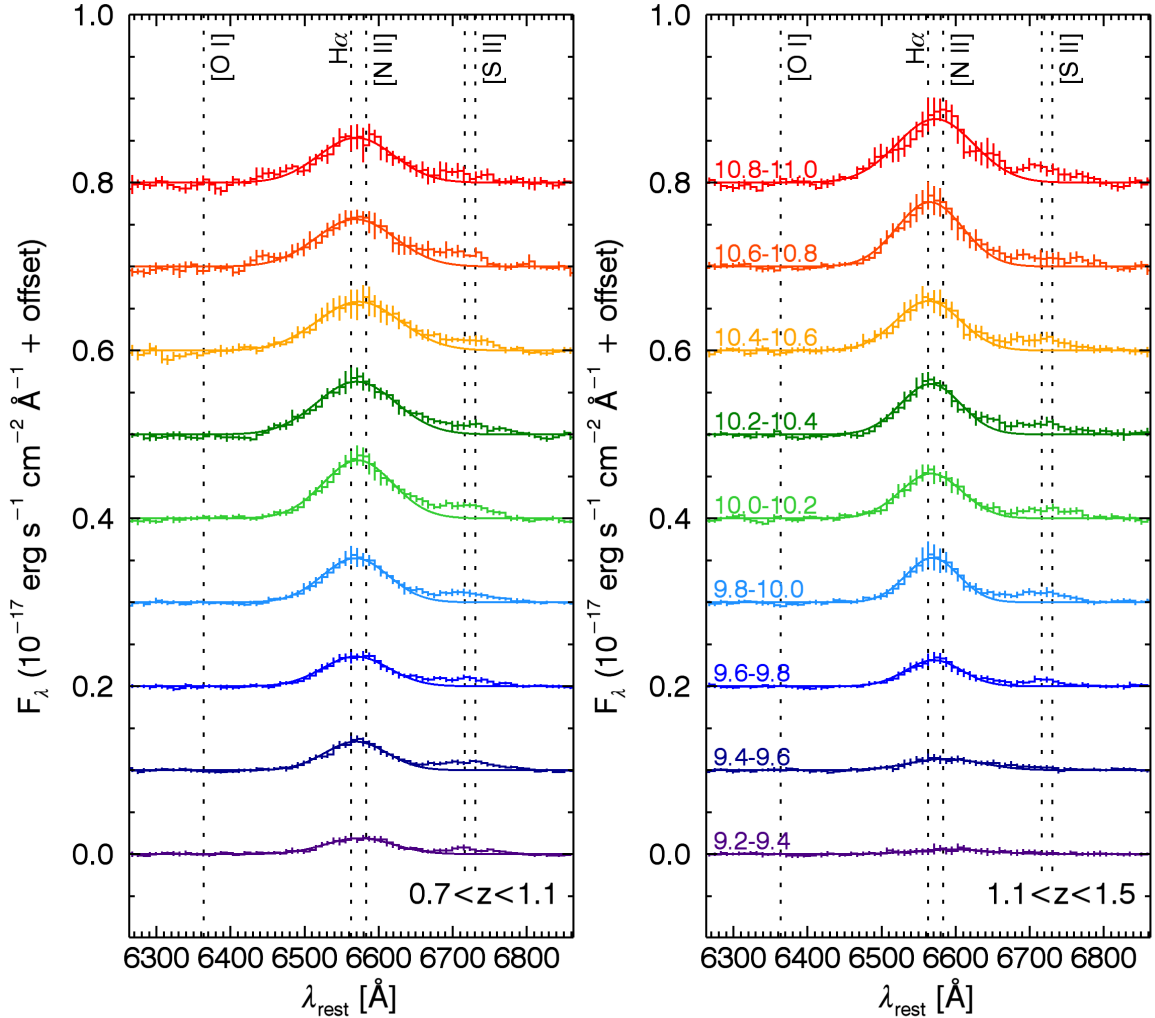}
\caption{H$\alpha$ stacks of HST/WFC3 G141 grism spectroscopy from the 3D-HST survey for all star-forming galaxies with coverage
at $0.7<z<1.5$, split in two redshift bins and nine bins of stellar mass with 0.2 dex width.  Where a grism redshift has
not yet been calculated (for $H_{F140W}>23$), we adopt the photometric redshift.  We measure the
H$\alpha$ emission line fluxes to derive SFRs. }
\label{fig:Halpha}
\end{figure*}

The use of the rest-frame UV continuum slope $\beta$ as a dust correction to L$_{\mathrm{UV}}$ remains largely 
uncertain with little consensus in the literature.  
Figure~\ref{fig:IRX_beta} shows correlation between IRX and $\beta$ from this work, 
compared to a range of best-fit relations from the literature.
The \citet{Meurer99} relation is derived from the UV spectra and fluxes of local starbursts and used to 
empirically calibrate the $\mathrm{L_{IR}/L_{UV}}$ correlation with $\beta$ in terms of the dust absorption at 1600$\mathrm{\AA}$.
This only agrees well with our data at $z>1.5$ and $\beta>-0.5$, whereas the data are lower by 0.2--0.5 dex otherwise.
It is difficult to say if the offset in the lowest redshift bin is intrinsic, as the rest-frame UV continuum is often sampled by only the two bluest photometric bands.  
The offset may also be due to systematic uncertainties in the 24$\mu$m to
total bolometric IR luminosity calibration as discussed in Section~\ref{sec:sysunc}, 
or some other unknown calibration error.

Recent work by \citet{Takeuchi12} finds a significantly lower $z=0$ relation for the same sample of local starburst galaxies
as \citet{Meurer99},
when aperture corrections to the UV flux are properly taken into account.
Our measurements of the IRX correlation with $\beta$ are identical to the \citet{Takeuchi12} $z=0$ relation at $0.5<z<1.0$,
and 0.1--0.4 dex higher at all other redshifts.
Taken at face value, the dust properties of galaxies at $z>1$ may be different from local star-forming galaxies.  
Similarly, \citet{Oteo14} find that the IRX-$\beta$ relations for local starbursts are
insufficient at high redshifts to recover the necessary dust-correction factors to reconcile the observed FIR with
the derived SFR.  They suggest an evolution of the dust properties of star-forming galaxies with redshift.  
We do not find evidence for a strong evolution in the average dust properties of galaxies at $0.5<z<2.5$, although
galaxies at $z>1$ do appear to have different dust properties than local starburst galaxies \citep[see also, e.g.,][]{Kashino13,Price14}.  
The strong redshift evolution observed by \citet{Oteo14} may be a result of their sample selection.

In another high redshift study, \citet{Heinis13} perform a stacking analysis of a sample of UV-selected galaxies at $z\sim1.5$
and find a correlation that underestimates the correlation derived from local starburst galaxies 
but is in good agreement with that derived from local normal star-forming galaxies.  
The correlation we measure at $z\sim1.5$ generally agrees with the \citet{Heinis13} relation.  
Galaxies at $z\sim1.5$ may therefore be more similar in their dust 
properties to ``normal'' local galaxies rather than starburst galaxies.  From Figure~\ref{fig:IRX_beta}, we see that there 
is little consensus amongst the measured correlations in the literature.  The relations presented in the literature show a larger
scatter than the self-consistent measurements presented herein. 

\begin{figure}[t]
\leavevmode
\centering
\includegraphics[width=\linewidth]{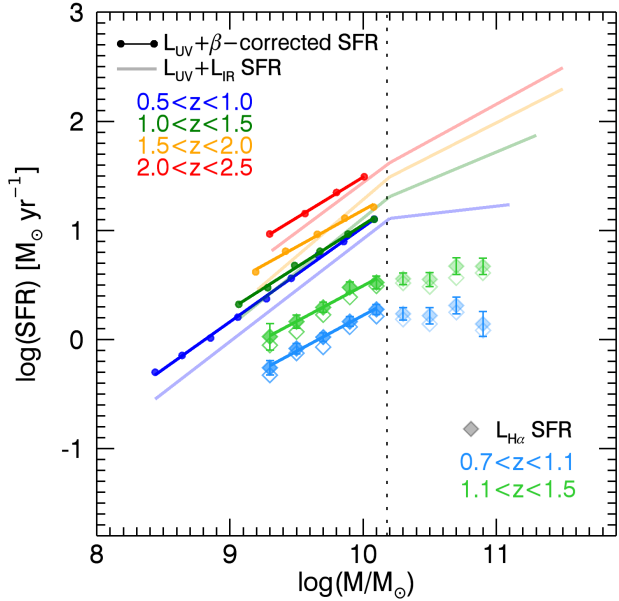}               
\caption{Comparison of the star formation sequence as measured with UV+IR SFRs (shown in transparency), relative 
to other SFR indicators independent of the FIR: H$\alpha$ SFRs (diamonds)
and UV luminosities corrected by the slope $\beta$ of the UV continuum.  The slope measured at low stellar masses is consistent amongst the 3
independent SFR indicators with a constant value of order $\alpha\sim0.8-1$ from $z=0.5$ to $z=2.5$.}
\label{fig:HalphaSFR}
\end{figure}

To estimate the total SFR from the UV alone, we correct $\mathrm{L_{UV}}$ for each bin in stellar mass by the average measured UV continuum slope,
$\mathrm{L_{2800,corr}}=\mathrm{L}_{2800}\times10^{0.4A_{2800}}$.  We use the \citet{Meurer99} relation, $\mathrm{A_{1600}}=4.43+1.99\beta$,
and the \citet{Calzetti00} dust law to convert $\mathrm{A_{1600}}$ to $\mathrm{A_{2800}}$.
We only consider stellar mass bins where $\beta<-0.5$, as \citet{Wuyts11a} show that dust-corrected SFR measurements above these limits become 
highly uncertain because the majority of the rest-frame UV light is absorbed (see also Figures~\ref{fig:LIR_LUV}
and \ref{fig:IRX_beta} here).  In practice, this limits us to low stellar masses where galaxies have less dust.  We discuss these measurements 
of the $\log\Psi-\log\mathrm{M_{\star}}$ relation in the following section, together with H$\alpha$ SFRs.

\subsection{H$\alpha$ Star Formation Rates}

The main uncertainty when inferring the total SFRs without FIR data remains the dust correction,
which becomes increasingly significant at shorter wavelengths.  Although the dust correction at 6500$\mathrm{\AA}$ is still significant, 
H$\alpha$ is less affected by dust than the UV continuum.  The 3D-HST G141 grism spectroscopy covers the H$\alpha$ 
emission line at $0.7<z<1.5$.  We therefore select all galaxies in this redshift range with 3D-HST coverage (which
includes roughly $\sim75\%$ of the CANDELS $J_{F125W}$ and $H_{F160W}$ imaging) and perform a stacking analysis to measure H$\alpha$
line fluxes.  Grism redshifts and emission line fluxes are
only measured down to $H_{F160W}=23$ mag for the 3D-HST v4.0 internal release\footnote{The public
release of the 3D-HST grism spectroscopy will derive grism redshifts and measure emission line fluxes for all objects},
whereas the one dimensional (1D) spectra are extracted for all objects.
We select galaxies with a S/N ratio greater than 10 in the $H_{F140W}$ direct image for two redshift bins at 
$0.7<z<1.1$ and $1.1<z<1.5$ and split the sample into the same 0.2 dex stellar mass bins as the UV+IR analysis,
using the photometric redshift where a grism redshift is not available for $H_{F160W}>23$. 

The 1D spectra are scaled to match the well-calibrated HST/WFC3 photometry by taking the ratio of $H_{F140W}$ to 
the average continuum flux in the G141 spectrum. We shift the spectra to the rest-frame and interpolate to a wavelength
grid with $\Delta\lambda=8\mathrm{\AA}$.  We fit a second order polynomial to the individual spectra, masking out the H$\alpha$ emission line, 
to parameterize and subtract the continuum.  For each stellar mass, we measure a median and average rest-frame spectrum from
the stacks.  We show the average H$\alpha$ stacks in Figure~\ref{fig:Halpha}.  In addition to H$\alpha$, we clearly detect the 
[SII] emission lines.  The S/N ratio in the G141 spectra becomes too below
$\log(\mathrm{M_{\star}/M_{\odot}})=9.2$ to make robust H$\alpha$ emission line flux measurements.   
The lower photometric redshift accuracy will act to broaden the emission lines, but not to the degree where we expect
H$\alpha$ to blend with [SII].  The error bars presented in Figure~\ref{fig:Halpha} are derived from 50 iterations of a 
bootstrap analysis.

Both the mean and median spectra are fit with a Gaussian (solid lines in Figure~\ref{fig:Halpha}), from which we 
measure the H$\alpha$ emission line fluxes.  Following \citet{Wuyts13}, we assume that [NII] contributes 15\% of the
measured H$\alpha$ emission line flux such that the [NII]/(H$\alpha$+[NII]) ratio equals 0.15.  
The SFR is derived from the H$\alpha$ flux using the conversion presented in \citet{Kennicutt98}, 
adapted from a Salpeter IMF to a \citet{Chabrier} IMF following \citep{Muzzin10}:

\begin{equation}
\Psi_{\mathrm{H\alpha}}~[\mathrm{M_{\odot}~yr^{-1}}]=1.7\times10^{-8}~\mathrm{L_{H\alpha}}~[\mathrm{L_{\odot}}].
\end{equation}

The measured H$\alpha$ SFRs are presented together with the SFRs derived from the $\beta$-corrected $\mathrm{L_{UV}}$ in 
Figure~\ref{fig:HalphaSFR}.  Filled circles and diamonds signify the UV+$\beta$-correction SFRs and the H$\alpha$ SFRs derived from the mean stacks, 
respectively, whereas
the open diamonds correspond to the measurements from median stacks.  
The $\beta$-corrected UV SFRs agree within 0.1 dex with the UV+IR SFRs at $\log\mathrm{M_{\star}}=10$ M$_{\odot}$, but become
increasingly higher by up to $\sim0.2$ dex at $\log\mathrm{M_{\star}}=9$ M$_{\odot}$.  We suspect these offsets are related to 
the large uncertainties in the conversion from $\beta$ to $A_{1600}$.
The H$\alpha$ SFRs are offset by $\sim0.5$ dex from the UV+IR SFRs, presumably because they 
do not include a dust correction.  This offset implies $\sim1.3$ magnitudes of extinction in H$\alpha$ for $\log(\mathrm{M_{\star}/M_{\odot}})<10.2$.
This value is larger than the estimates of \citet{Garn10} that range from $\mathrm{A_{H\alpha}}=0.3-1$ at $\log\mathrm{M_{\star}}\sim9-10$ M$_{\odot}$.
Although stacks of emission-line detected galaxies with grism redshifts are consistent with the running mean of the individual measurements, 
we cannot rule out uncertainties in our analysis introduced due to our use of photometric redshifts for the larger (lower-mass) sample.
Above $\log(\mathrm{M_{\star}/M_{\odot}})=10.2$, the H$\alpha$
SFRs are almost flat, which we interpret as due to the large dust corrections necessary at the massive end.
Due to the strong correlation between the amount of dust in a galaxy (as probed through
the IRX ratio) and the stellar mass, dust corrections will only act to steepen the derived $\log\Psi-\log\mathrm{M_{\star}}$ 
relation.  The slope we measure from the H$\alpha$ SFRs for $\log(\mathrm{M_{\star}/M_{\odot}})<10.2$ therefore acts as a lower limit,
independent of the other SFR indicators explored in this paper.  

\begin{figure}[t]
\leavevmode
\centering
\includegraphics[width=\linewidth]{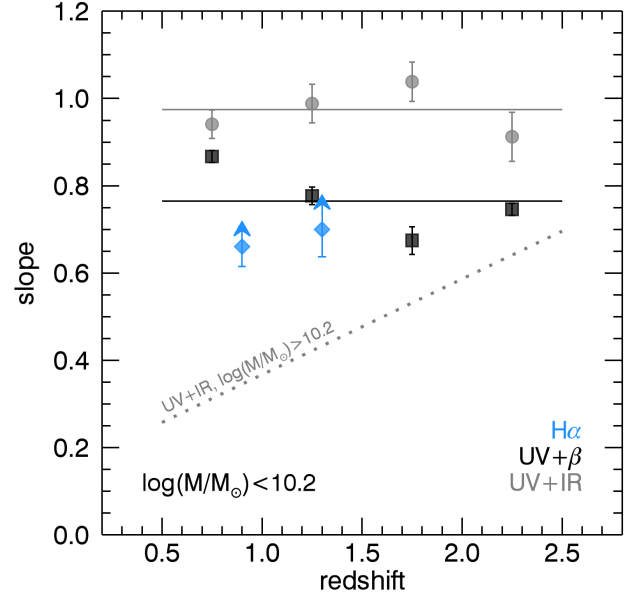}
\caption{Evolution of the low mass slope ($\log\mathrm{M_{\star}}<10.2$ M$_{\odot}$) with redshift for SFR indicators independent of the IR.
The low mass slope of the $\log\Psi-\log\mathrm{M_{\star}}$ relation is roughly constant from $z=0.5$ to $z=2.5$, as measured from 3 
independent SFR indicators.  The H$\alpha$ SFRs do not include any dust correction
and therefore serve as lower limits to the slope due to the strong correlation between the amount of dust in a galaxy and its
stellar mass.  The solid horizontal lines represent the average measured low mass slopes for the different SFR indicators. }
\label{fig:betaslope}
\end{figure}

\begin{figure*}[t]
\leavevmode
\centering
\plottwo{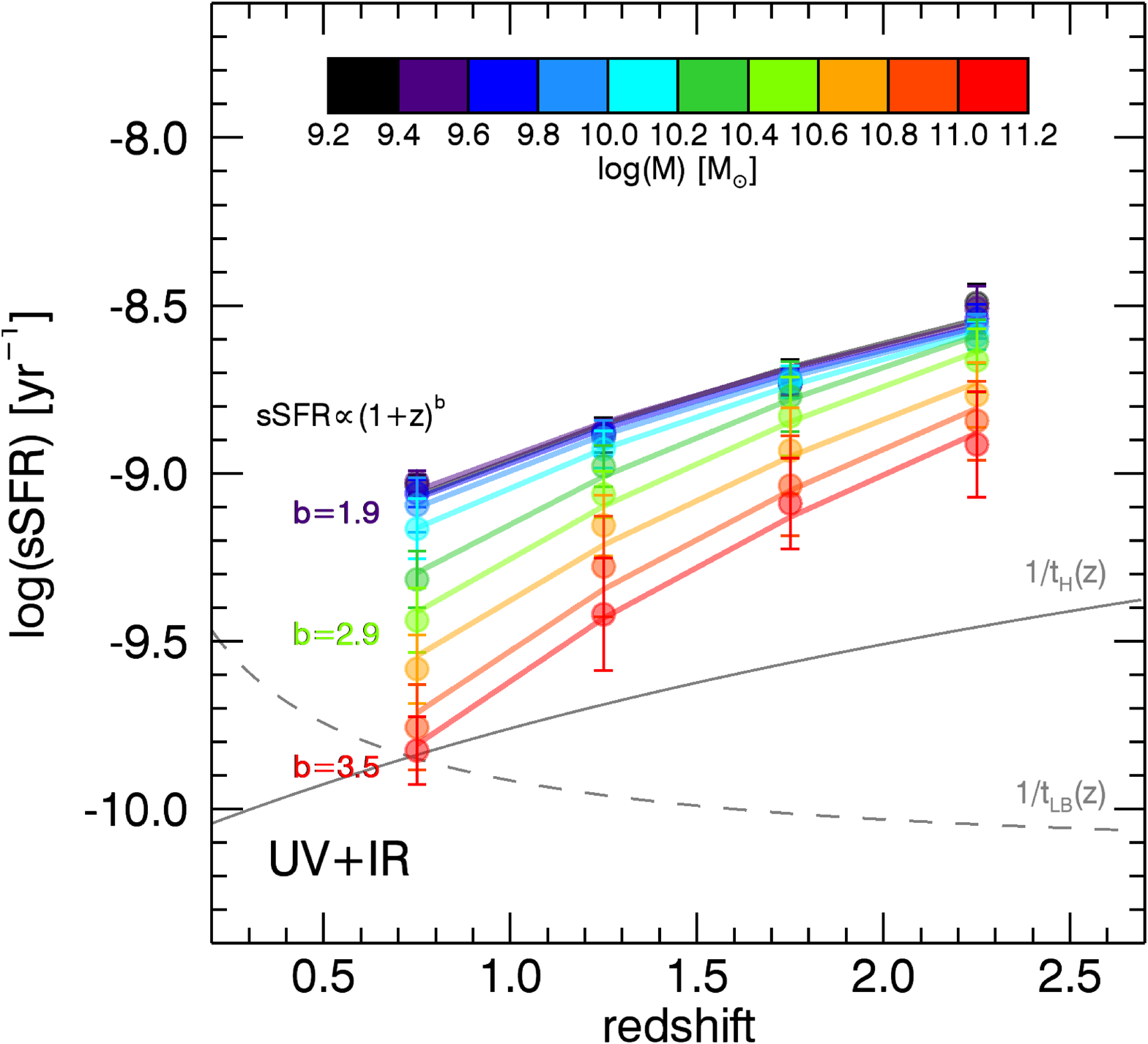}{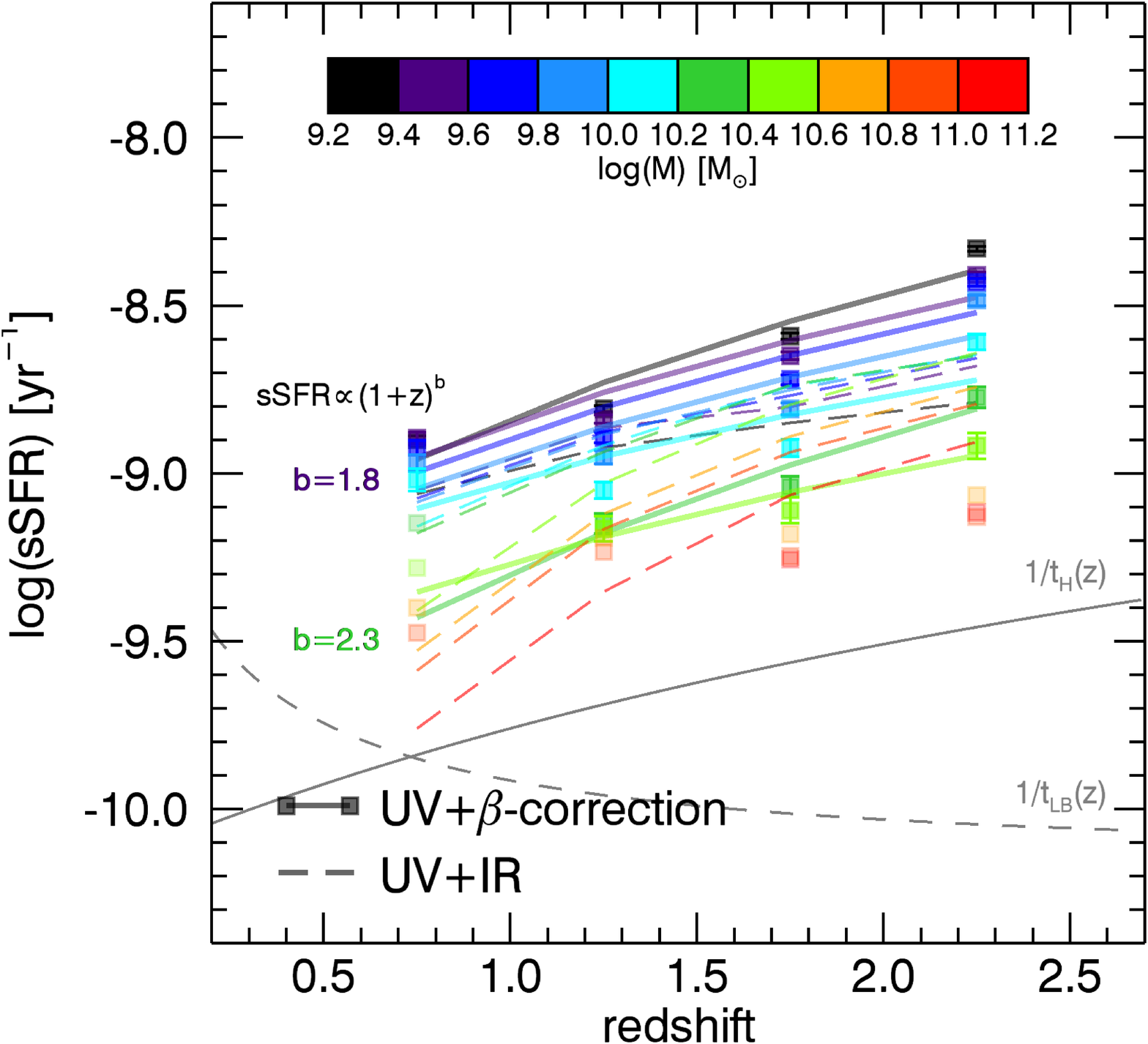}
\caption{Evolution of the sSFR with redshift in bins of stellar mass.
The sSFR exhibits a mass-dependent redshift evolution in both the UV+IR SFRs (left) and the
$\beta$-corrected UV SFRs (right).  Whereas the sSFRs of the most massive galaxies evolve strongly with
redshift, all galaxies less massive than $\log(\mathrm{M_{\star}/M_{\odot}})=10$ show roughly the same shallower redshift
evolution. The $\beta$-corrected UV SFRs exhibit a slightly shallower redshift evolution than the UV+IR SFRs,
however large uncertainties remain in the $\beta$ calibration.}
\label{fig:ssfr_z}
\end{figure*}

We compile the measurements of the slope of the $\log\Psi-\log\mathrm{M_{\star}}$ relation below $\log(\mathrm{M_{\star}/M_{\odot}})=10.2$
in Figure~\ref{fig:betaslope}.  Both the UV continuum and H$\alpha$ measurements are consistent with a steeper slope for low mass galaxies
at all redshifts.  Both the UV+IR and UV continuum low-mass slopes are constant with
redshift, but systematically offset by 0.2.  These differences may be due to the large
uncertainties remaining in the $\beta$-$A_{1600}$ calibration, as evident from Figure~\ref{fig:IRX_beta}.  

\section{Evolution in the Specific Star Formation Rate}
\label{sec:ssfr}

The mass-dependent time evolution of the sSFR is presented in Figure~\ref{fig:ssfr_z} for the UV+IR SFRs (left panel) and the
SFRs derived from the $\beta$-corrected UV luminosity (right panel).  Following previous studies in the literature \citep[e.g.,][]{Damen09},
we parameterize the redshift evolution of the sSFR as,

\begin{equation}
\frac{\Psi}{\mathrm{M_{\star}}}~[\mathrm{yr^{-1}}] = a(1+z)^{b}.
\label{eq:ssfr}
\end{equation}

We present the best-fit values for both the UV+IR and $\beta$-corrected UV sSFRs in Table~\ref{tab:ssfr}. We observe
a mass-dependent redshift evolution in both SFR indicators, where galaxies less massive than $\log(\mathrm{M_{\star}/M_{\odot}})=10$
exhibit a similar evolution with $b=1.9$ for UV+IR SFRs and a slightly lower average value of $b=1.8$ for $\beta$-corrected UV SFRs.
Galaxies more massive than $\log(\mathrm{M_{\star}/M_{\odot}})=10$ show a stronger redshift evolution, reaching a maximum
value of $b=3.5$ for $\log(\mathrm{M_{\star}/M_{\odot}})>11$.  As discussed in Section~\ref{sec:UVcontinuum}, 
shallow UV continuum slopes are not well-calibrated and we therefore 
cannot accurately measure the redshift evolution from the $\beta$-corrected UV SFRs for galaxies more massive than 
$\log(\mathrm{M_{\star}/M_{\odot}})=10.5$.  

A plausible explanation for the decline in the sSFR since $z\sim2$ is a decrease in the gas accretion rate on to galaxies \citep[e.g.,][]{Dutton10}.
The specific accretion rate for dark matter haloes scales as $(1+z)^{2.25}$ for $z<2$ \citep{Birnboim07}, similar to the average evolution
in the sSFR for low-mass galaxies.  Semi-analytic models (SAMs) predict a similar evolution with $b=2.5$ \citep{QGuo11}, whereas
the observed redshift evolution from previous studies is larger with $b\sim3-5$ \citep[e.g.,][]{Salim07,Damen09,Karim11},
and only marginally consistent with the evolution we measure for the most massive galaxies.
The redshift evolution we measure for low mass galaxies agrees to first order with the theoretical predictions, whereas the 
more rapid evolution for massive galaxies remains discrepant with models.  This discrepancy likely relates to the poorly understood
physics of the quenching of star-formation. 

\begin{table*}[t]
\centering
\begin{threeparttable}
    \caption{Evolution of the specific star formation rate}\label{tab:mass}
    \begin{tabular}{lcccc}
      \hline \hline
      & \multicolumn{2}{c}{UV+IR} & \multicolumn{2}{c}{UV+$\beta$} \\
      \cline{2-3} \cline{4-5}
      $\log(\mathrm{M_{\star}/M_{\odot}})$~~~~~~ &  ~~~~~~~~~$\log a$~~~~~~~~~ & ~~~~~~~~~$b$~~~~~~~~~ & ~~~~~~~~~$\log a$~~~~~~~~~ & ~~~~~~~~~$b$~~~~~~~~~ \\
      \hline
      \noalign{\smallskip}
      9.2--9.4 & $ -9.54\pm0.15$ & $1.95\pm0.24$  & $-9.47\pm0.23$ & $2.10\pm0.37$ \\
      9.4--9.6 & $ -9.50\pm0.14$ & $1.86\pm0.22$ & $ -9.39\pm0.23$ & $1.78\pm0.37$ \\
      9.6--9.8 & $ -9.54\pm0.07$ & $1.90\pm0.12$ & $ -9.43\pm0.30$ & $1.77\pm0.48$ \\
      9.8--10.0 & $ -9.58\pm0.03$ & $1.98\pm0.04$ & $ -9.46\pm0.34$ & $1.70\pm0.54$ \\
      10.0--10.2 &  $ -9.69\pm0.03$ & $2.16\pm0.04$ & $ -9.45\pm0.37$ & $1.43\pm0.59$ \\
      10.2--10.4 & $ -9.93\pm0.08$ & $2.63\pm0.12$ & $ -9.99\pm0.34$ & $2.31\pm0.52$ \\
      10.4--10.6 & $-10.11\pm0.10$ & $2.88\pm0.16$ & $ -9.72\pm0.29$ & $1.52\pm0.43$ \\
      10.6--10.8 & $-10.28\pm0.15$ & $3.03\pm0.24$ & -- & -- \\
      10.8--11.0 & $-10.53\pm0.17$ & $3.37\pm0.26$ & -- & --  \\
      11.0--11.2 & $-10.65\pm0.11$ & $3.45\pm0.17$ & -- & --  \\
      \noalign{\smallskip}
      \hline
      \noalign{\smallskip}
    \end{tabular}
    \begin{tablenotes}
      \small
    \item \emph{Notes.} Redshift evolution of the sSFR is parameterized in Equation~\ref{eq:ssfr}.
      We include the redshift evolution as measured from the UV+IR SFR indicator and the $\beta$-corrected UV SFR.
      The $\beta$-corrected UV SFR is unreliable at the highest stellar masses and we therefore only include measurements
      for $\beta<-0.5$.
    \end{tablenotes}
    \label{tab:ssfr}
  \end{threeparttable}
\end{table*}

\section{Broader Implications}
\label{sec:broad}

In this paper, we present an empirical study of the properties of the star formation sequence over roughly half of cosmic time ($0.5<z<2.5$).
The three main observable quantities of the star formation relation, the normalization, intrinsic scatter, and slope, encapsulate 
fundamental physical quantities that regulate star formation.
The normalization of the star formation sequence is governed predominantly by the changing cosmological gas accretion rates with redshift.
The intrinsic scatter of this relation reveals the level of stochasticity in the gas accretion history. Lastly, the measured slope of
this relation tells about star formation efficiency.  As the various different feedback mechanisms are known to dominate
at different stellar mass regimes, it should perhaps come as no surprise that we now measure for the first time different slope and
normalization evolution for low and high mass galaxies.

We measure the slope of the star formation sequence for complete samples of both low mass
and high mass galaxies at $0.5<z<2.5$, finding a steeper slope for less massive galaxies.  From a radio
stacking analysis, \citet{Karim11} found tentative evidence for curvature of the star formation 
sequence.  However, they only considered blue star-forming galaxies and all the deviations
occurred below the mass representativeness limits, so no conclusions could be reached.  Similarly, \citet{Whitaker12b}
explored the changing slope of the star formation sequence in greater detail at $z\sim1$, finding that the bluest
(lowest mass) galaxies had a steeper slope than redder (high mass) galaxies.  As the color of a galaxy is directly 
correlated with the amount of dust, these observations demonstrate why measurements
of the slope from UV SFRs are often closer to unity ($\alpha\sim0.75-1$), whereas IR SFRs are shallower with a slope of $\alpha\sim0.6$ \citep{Speagle14}.
Here, we present the first 
statistically robust measurement of the low-mass slope of the star formation sequence using UV+IR SFRs for mass-complete 
samples at $0.5<z<2.5$.

We observe that the star formation sequence has a roughly constant slope of $\alpha\sim1$ for $\log(\mathrm{M_{\star}/M_{\odot}})<10.2$.
Dust-corrected H$\alpha$ measurements of the star formation sequence for both star-forming and quiescent galaxies in the 
SDSS by \citet{Brinchmann04} result
in a slightly shallower slope of $\alpha=0.7$, with a drop in SFR beyond $\log(\mathrm{M_{\star}/M_{\odot}})=10$.
Local studies by \citet{Huang12} suggest a transition mass of $\log(\mathrm{M_{\star}/M_{\odot}})\sim9.5$ below 
which star formation scales differently with total stellar mass \citep[also found by][]{Salim07, Gilbank11}, with a steeper slope.  
\citet{Kannappan09} identify a similar threshold stellar 
mass below which the number of blue sequence galaxies with elliptical morphologies sharply rises. 

\begin{figure}[t]
\leavevmode
\centering
\includegraphics[width=\linewidth]{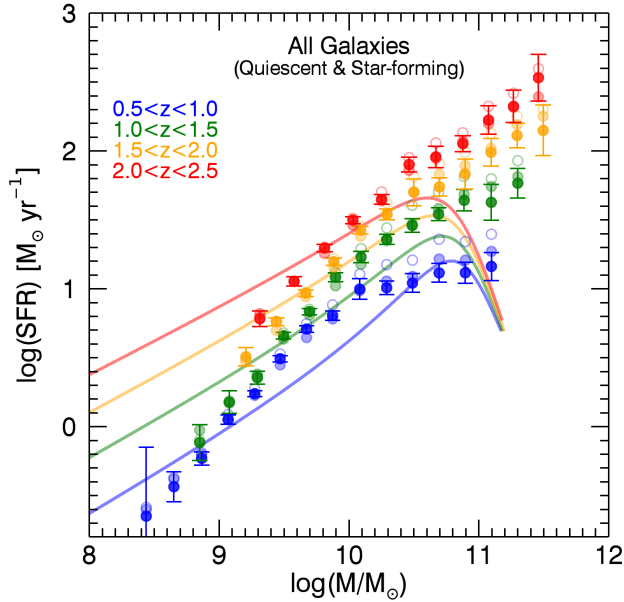}
\caption{Comparison between the UV+IR SFRs derived from a MIPS/24$\mu$m stacking analysis of all galaxies
and predictions from the \citet{Dave12} analytical equilibrium models,
assuming momentum-driven winds and a quenching mass of $\log(\mathrm{M_{halo}/M_{\odot}})=12.3$.  The UV+IR SFRs
from the average (light open circles) and median (light filled circles) stacking analyses for
star-forming galaxies only are shown for reference.
These simple equilibrium
models fail to match the steep slope at low masses, as well as the overall normalization, suggesting either non-equilibrium
conditions or a departure from self-similar prescriptions for feedback.}
\label{fig:dave}
\end{figure}

Many theoretical models predict a steep relation between $\log\mathrm{M_{\star}}$ and $\log\Psi$, 
but have trouble producing the shallow slope found for the massive star-forming galaxies \citep[e.g.,][]{Somerville08}.
The fact that we find a steep slope for the low-mass galaxies will make it easier to reconcile the galaxy 
formation models with the observations.  The simple equilibrium models from \citet{Dave12} are a notable exception, they 
actually predict a rather shallow slope at all masses, inconsistent with the results obtained here.  
Figure~\ref{fig:dave} presents a comparison between the momentum-driven wind models of the \citet{Dave12} analytical framework, as outlined in detail
in Section 7.1 of \citet{Henry13a}, and UV+IR SFRs measured from stacks of all galaxies (see Appendix C). 
Other studies favor these simple models because they produce the best-fit to 
a variety of quasar absorption line studies of the intergalactic medium, while
also reproducing the observed cosmic star formation history \citep[e.g.,][]{Oppenheimer06,Oppenheimer08}.  
There are two possible explanations for the differences between the predictions from the equilibrium model of \citet{Dave12} and other 
models. It could either be that the low-mass systems are not in equilibrium, as implicitly assumed in the \citet{Dave12} framework, 
and the time-dependence of feedback must be properly quantified.  Or it may be that the wind-recycling in these low-mass 
systems has been underestimated \citep[e.g.,][]{Oppenheimer10}.  Qualitatively, non-equilibrium models match the observed
mass and redshift evolution of the SFR \citep[e.g.,][see also summary in Figure 9 of Leja et al. 2014]{KGuo13,Mitchell14,Behroozi14,Genel14}.
Future detailed comparisons between the observed properties of the star formation sequence and models may 
place tighter constraints on the various feedback processes that galaxies undergo.

In a recent paper, Leja et al. (2014) examine the connection between the observed correlation between
$\log\Psi$ and $\log\mathrm{M_{\star}}$ from \citet{Whitaker12b} and the observed stellar mass
functions from \citet{Tomczak14} at $0.2<z<2.5$.  They find that an extrapolation of
the relatively shallow slopes measured by \citet{Whitaker12b} cannot hold true at the low-mass end,
as this results in an extremely rapid growth of the stellar mass function at low stellar masses that is not observed.
With a completely independent analysis, we show here that the slope for less massive galaxies is steeper
than that measured for more massive galaxies.  The mass function analysis by Leja et al. (2014) predicts a
low-mass slope of $\alpha=0.9-1.1$, in agreement with
the UV+IR SFRs, but slightly steeper than the $\beta$-corrected UV SFRs.
Leja et al. (2014) note that some discrepancies between the
stellar mass function and star formation sequence remain, even after adopting a low-mass slope of unity.
For example, the normalization of the SFRs is $\sim0.3$
dex higher than predicted from the growth of the mass function at $1<z<2.5$ \citep[Leja et al. 2014; see also][]{Genel14}.

We find a strong evolution in the slope of the star formation sequence for galaxies more massive than $\log(\mathrm{M_{\star}/M_{\odot}})>10.2$. 
This redshift evolution in the high-mass slope, combined with the redshift-dependent low-mass limit of the NMBS analysis,
is likely what drove the evolution of the derived slope in \citet{Whitaker12b}.
This strong evolution may be related to the growth 
of the bulge.  \citet{Abramson14} find that when accounting for the bulge/disk decomposition, the SFR renormalized by the 
disk stellar mass reduces the stellar mass dependence of star formation efficiency by $\sim0.25$ dex per dex, reducing the slope by 0.25.
Here we find that the difference between the slope for low-mass galaxies and high-mass galaxies increases from $\Delta\alpha=0.3$ at $z\sim2$ 
to $\Delta\alpha=0.7$ at $z\sim0.7$.  
Indeed, \citet{Nelson12} find evidence for the rapid formation of compact bulges and large disks at $z\sim1$.
Similarly, \citet{Lang14} observe that the bulge to disk ratio increases by a factor of two from $\log(\mathrm{M_{\star}/M_{\odot}})=10$ to 
$\log(\mathrm{M_{\star}/M_{\odot}})=11.5$.  However, if the evolution of the high-mass slope is driven entirely by the growing
contribution from the bulge to the stellar mass, \citet{Lang14} should have also measured redshift evolution in the correlation 
between the bulge-to-disk and stellar mass.  The average bulge-to-disk ratio is consistent between their two
redshift bins at $z\sim1$ and $z\sim2$. 
We further caution that the formation of today's bulge-disk systems was a complex process; in particular, galaxies with 
the present-day mass of the Milky Way increased their mass in the central regions at only a slightly smaller rate than 
at $r>2$ kpc \citep[e.g.,][]{vanDokkum13,Patel13}. Furthermore, as galaxies grow in mass, the sSFRs of individual 
galaxies likely evolved in a different way than the mean sSFR at fixed mass \citep[e.g.,][]{Fumagalli12,Leja13}.
The idea that the shallower slopes measured for more massive galaxies is telling 
us about the growth of bulges in galaxies is intriguing, but warrants further analysis. 

\begin{figure}[t]
\leavevmode
\centering
\includegraphics[width=\linewidth]{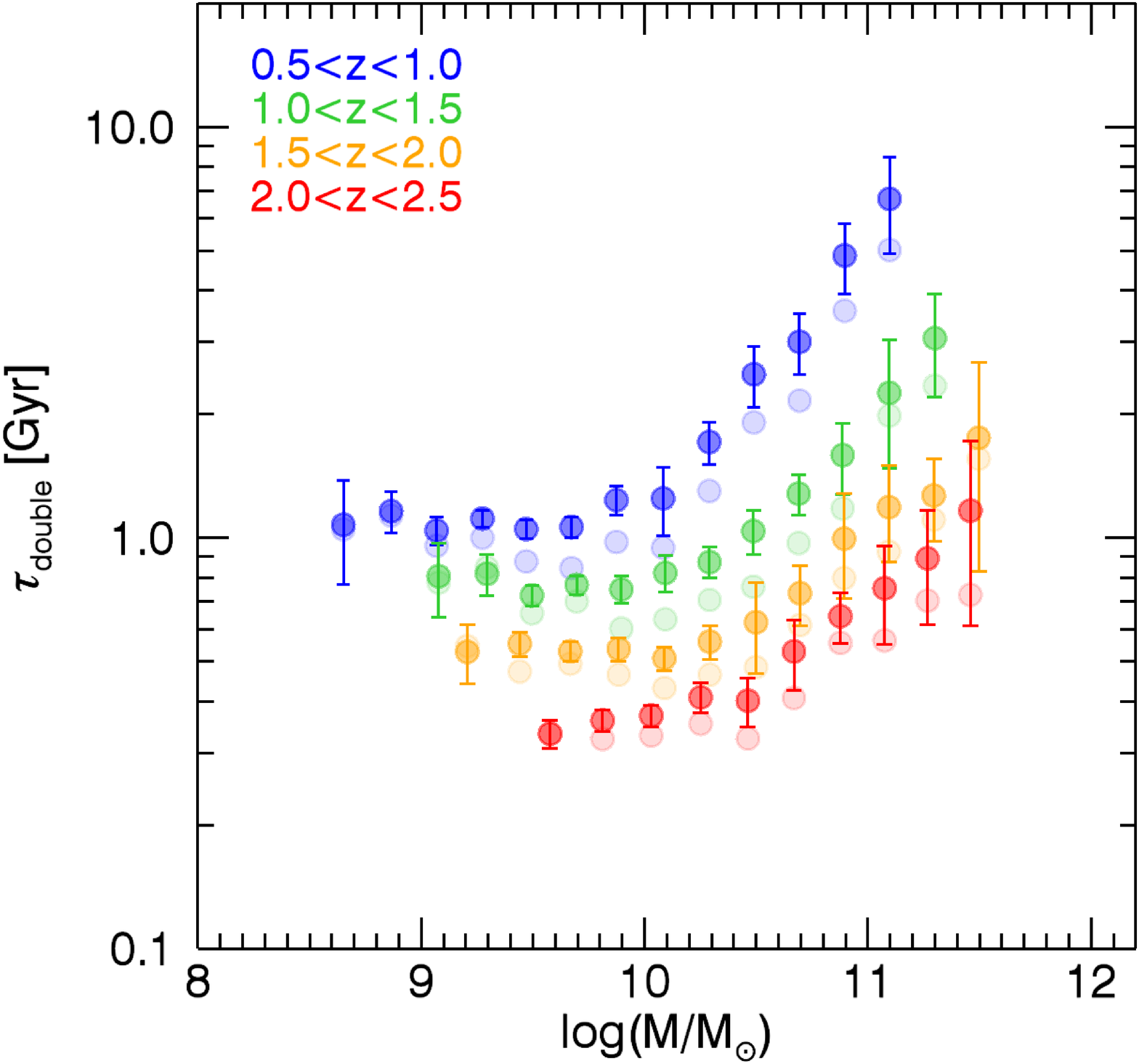}
\caption{The mass-doubling timescale (sSFR$^{-1}$) as a function of stellar mass for star-forming galaxies.
The filled circles with lighter colors represent the average stacks, whereas the darker circles represent median stacks.
The highest stellar mass galaxies at a given epoch exhibit
the longest mass-doubling timescales, whereas this timescale increases towards lower redshifts at fixed stellar
mass. The mass-doubling timescale of low-mass galaxies evolves in a roughly self-similar fashion. }
\label{fig:massdouble}
\end{figure}

In Figure~\ref{fig:massdouble}, we present the mass-doubling timescale.
The inverse of the sSFR of a galaxy is often interpreted as a physical timescale for the formation of the stellar population,
where $(\Psi/\mathrm{M_{\star}})^{-1}$ is equivalent to the time it would take for the stellar mass of a galaxy to double.
We include both the median and average stacks in Figure~\ref{fig:massdouble}, as denoted by the darker and lighter colors, respectively.
The mass-doubling timescale is roughly self-similar for low-mass galaxies, evolving towards shorter formation timescales at earlier times.
Whereas galaxies at $z\sim0.5$ will double their mass in about a gigayear, this timescale is only 300 Myr at $z\sim2$.  This
stellar mass doubling timescale implies high gas accretion rates at earlier times.
The observed trends in the timescale for a galaxy to double
in mass sets interesting constraints on feedback prescriptions in future efforts for galaxy formation theories.

Here, we explore the average properties of the star formation sequence a factor of ten lower in stellar mass than previous studies
for a single SFR indicator.  With this unique dataset, we are therefore able to study the mass-dependencies of the normalization in
addition to the slope.
Previous studies of the redshift evolution of the sSFR (sensitive to the normalization of the star formation sequence) 
find results similar to our measurements for the most massive 
galaxies \citep[e.g.][]{Salim07,Karim11}, with $\Psi/\mathrm{M_{\star}}\propto(1+z)^{3-4}$.  We demonstrate in this paper
that less massive galaxies exhibit a more gradual redshift evolution of their sSFRs, similar to the evolution of 
the cosmological gas accretion rate.   
We observe the redshift evolution in the sSFR to be self-similar for galaxies
less massive than $\log(\mathrm{M_{\star}/M_{\odot}})=10$ with $\Psi/\mathrm{M_{\star}}\propto(1+z)^{1.9}$, whereas more massive galaxies show a
stronger redshift evolution with $\Psi/\mathrm{M_{\star}}\propto(1+z)^{2.2-3.5}$ for $\log(\mathrm{M_{\star}/M_{\odot}})=10.2-11.2$. 
The redshift evolution for less massive galaxies is slightly shallower when using total SFRs derived from rest-frame UV
luminosities corrected for dust by measuring the UV continuum shape.  To our knowledge,  such shallow redshift evolution in 
the sSFRs of galaxies has not been observed previously.  
Although \citet{Sobral14} don't see this strong mass dependence, their H$\alpha$ SFR measurements are highly dependent 
on the dust correction and their sample is not mass-selected.  
On the other hand, \citet{Behroozi13} do find a similarly strong mass-dependence for the redshift 
evolution of the sSFR by connecting galaxies across all different epochs via abundance matching to halos in dark matter 
simulations. Similarly, 
although the absolute normalization is offset, the trend for galaxies 
less massive than $\log(\mathrm{M_{\star}/M_{\odot}})=10$ to
show similar redshift evolution in their sSFRs qualitatively agrees with recent results from cosmological hydrodynamical simulations
\citep{Genel14}.  

\section{Conclusions}

In this paper, we combine the deep photometry available in the CANDELS legacy fields with the grism redshifts and H$\alpha$ emission line measurements
from the 3D-HST treasury program, together with Spitzer/MIPS 24$\mu$m imaging.
This unique combination of data allows us to leverage
the strengths of each respective survey and enables significant improvements to our understanding of the star formation sequence
relation across the full dynamic range in stellar masses over roughly half of cosmic history.
With a mass-complete sample of 39,106 star-forming galaxies (out of a total sample of 58,973 star-forming galaxies) selected from the
public 3D-HST photometric catalogs\footnote{\url{http://3dhst.research.yale.edu/Data.html}}, we provide the average measured UV+IR SFRs
from $z=0.5$ to $z=2.5$.  For the first time, we firmly distinguish between distinct high and low mass slopes.  The main results of our
analysis are:

\begin{itemize}
\item We find that low-mass galaxies with $\log(\mathrm{M_{\star}/M_{\odot}})<10.2$ evolve in a self-similar fashion with a constant slope
of unity ($\log\mathrm{\Psi}\propto\log\mathrm{M_{\star}}$), whereas
we observe a strong evolution in the slope for more massive galaxies ranging from $\log\mathrm{\Psi}\propto(0.3-0.6)\log\mathrm{M_{\star}}$ from 
$z=0.5$ to $z=2$.

\item We compare the total UV+IR SFRs, calibrated from a stacking analysis of Spitzer/MIPS 24$\mu$m imaging, to $\beta$-corrected 
UV SFRs and H$\alpha$ SFRs, finding that the low-mass slopes are consistently steeper than the high-mass slopes.

\item We confirm previous studies, showing that the average IRX ratio ($\mathrm{L_{IR}/L_{UV}}$, a proxy for the amount of dust in a galaxy) is
strongly correlated with stellar mass.  For the first time, we show that this average relation is consistent with no redshift evolution from
$z=0.5$ to $z=2.5$ for $9.0<\log\mathrm{M_{\star}}<10.5$ M$_{\odot}$.

\item The redshift evolution of the normalization varies in different mass regimes.  For galaxies
less massive than $\log(\mathrm{M_{\star}/M_{\odot}})<10$ the specific SFR ($\Psi/\mathrm{M_{\star}}$) is observed to be self-similar
with $\Psi/\mathrm{M_{\star}}\propto(1+z)^{1.9}$.  More massive galaxies show a
stronger evolution with $\Psi/\mathrm{M_{\star}}\propto(1+z)^{2.2-3.5}$ for $\log(\mathrm{M_{\star}/M_{\odot}})=10.2-11.2$.

\end{itemize}

Despite the many potential systematic uncertainties that remain in different SFR indicators, there
appears to be a consensus forming with regards to the redshift evolution and mass dependencies of star formation.  
It would be valuable to analyze longer wavelength imaging to estimate mid-IR fluxes for the present dataset; 
our current analysis is limited to 24$\mu$m. We note that \citet{Rodighiero14} analyzed Herschel 
160$\mu$m imaging, and found that 
the star formation rates estimated from stacks agreed well between star formation rates from 160$\mu$m, 
24$\mu$m, and dust-corrected UV luminosities.  Exploration of the properties
for the lowest mass galaxies beyond those presented in this paper is rich with potential for 
shaping our understanding of galaxy formation and evolution.

\begin{acknowledgements}
We thank the anonymous referee for insightful comments and a careful reading of the manuscript.
The authors wish to acknowledge Kristian Finlator for helpful discussions.
KEW and AH are supported by appointments to the NASA
Postdoctoral Program at the Goddard Space Flight Center,
administered by Oak Ridge Associated Universities through a
contract with NASA.
The authors are grateful to the many colleagues who have
provided public data and catalogs in the five deep 3D-HST/CANDELS fields; high redshift galaxy science
has thrived owing to this gracious mindset and the TACs and the Observatory Directors who have encouraged this.
This work is based on observations taken by the 3D-HST Treasury Program (GO 12177 and 12328)
with the NASA/ESA HST, which is operated by the Associations of Universities for Research in Astronomy, Inc.,
under NASA contract NAS5-26555.
\end{acknowledgements}

\appendix

\section{Emission Line Corrections to Photometry and Stellar Masses}\label{app:mass}

\begin{figure}[t]
\leavevmode
\centering
\includegraphics[width=\linewidth]{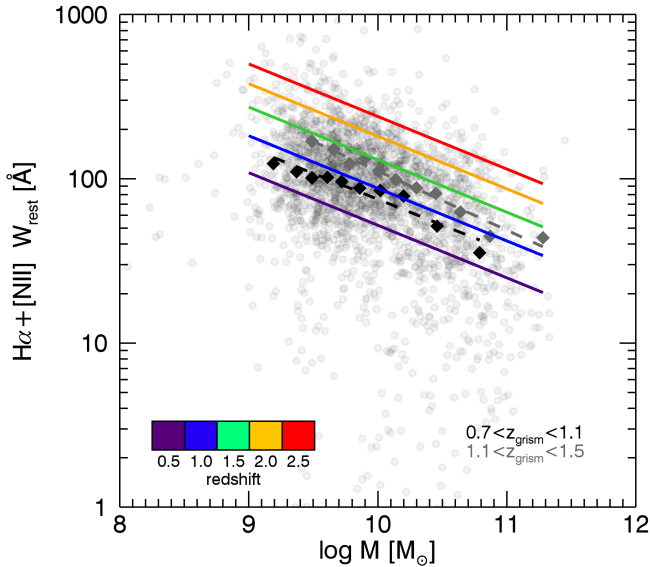}
\caption{Observed $\mathrm{H\alpha+N[II]}$ rest-frame equivalent widths of individual galaxies from the
3D-HST survey (circles) at $0.7<z<1.5$ are a strong function of stellar mass.  We combine the linear best-fits (dashed lines) to the
running means (diamonds) in two redshift bins with the redshift evolution from \citet{Fumagalli12} of $W\propto(1+z)^{1.8}$
to derive $W(z,\mathrm{M_{\star}})$ in Equation~\ref{eq:EW}.}
\label{fig:EW_indiv}
\end{figure}

\begin{figure*}[t]
\leavevmode
\centering
\includegraphics[width=0.75\linewidth]{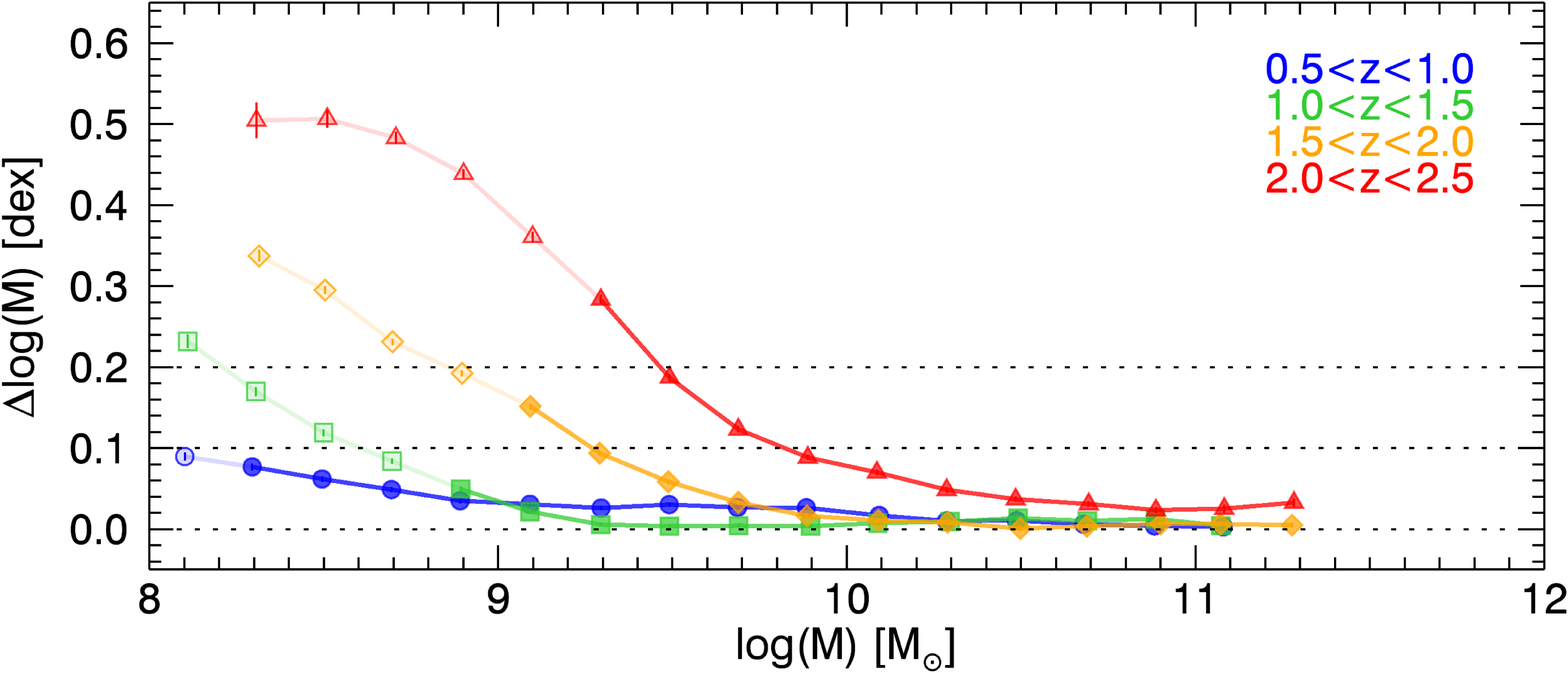}
\caption{The contamination to stellar mass estimates as a function of stellar mass and redshift. 
The typical contamination from emission line flux to the broadband and medium-band photometry becomes important only for
lower mass galaxies with $\log(\mathrm{M_{\star}/M_{\odot}})<9$, with the strongest trends observed at $2.0<z<2.5$.
We compare the original stellar masses to those derived from the
same stellar population synthesis modeling assumptions but including an additional subtraction of the contaminating emission line
flux in the relevant photometric bands.  On average, the stellar masses agree within $<0.04$ dex down to $\log(\mathrm{M_{\star}/M_{\odot}})=9.5$.}
\label{fig:comparemass}
\end{figure*}

To estimate the direct effect of the contamination from emission line flux to the estimated stellar masses, we
empirically derive the average correlation between the $\mathrm{H\alpha+N[II]}$ equivalent width ($W$) and stellar mass from individual 3D-HST
line flux measurements in two redshift bins, $\mathrm{0.7<z_{grism}<1.1}$ and $\mathrm{1.1<z_{grism}<1.5}$ in Figure~\ref{fig:EW_indiv}.
At the HST/WFC3 grism resolution, the H$\alpha$ line is blended with [NII].  For the purposes of determining the emission-line contamination,
it is not necessary to deblend these lines.
To account for the redshift evolution of the $W-\log\mathrm{M_{\star}}$ relation we adopt the slope from the best-fit relation
calculated by \citet{Fumagalli12}
of $W_\mathrm{{H\alpha}}\propto(1+z)^p$, where $p=1.8$.  We note that we are extrapolating the measurements by 
\citet{Fumagalli12} to lower stellar masses.  Combining the redshift evolution from \citet{Fumagalli12} with
the $\log W_\mathrm{{H\alpha+[NII]}} - \log\mathrm{M_{\star}}$ relation
we measure in Figure~\ref{fig:EW_indiv},
we parameterize the redshift and mass dependence of the rest-frame H$\alpha$+[NII] $W$ as,

\begin{equation}
\log W_\mathrm{{H\alpha}+[NII]} = 1.8\log(1+z)-0.32\left(\frac{\log\mathrm{M_{\star}/M_{\odot}}}{10^{10}}\right).
\label{eq:EW}
\end{equation}

In addition to the H$\alpha$+[NII] emission line flux contamination, we also estimate the contribution 
from the [OII]$\lambda$3727, 3729$\mathrm{\AA}$ and
[OIII]$\lambda$4959,5007$\mathrm{\AA}$ doublets and H$\beta$.  
From the individual 3D-HST line flux measurements, we calculate an average
$\log(F_{\mathrm{[OIII]}}/F_{\mathrm{H\beta}})$ value of $0.49\pm0.04$ at 
$\log(\mathrm{M_{\star}/M_{\odot}})=10.2$ and $0.43\pm0.05$ at $\log(\mathrm{M_{\star}/M_{\odot}})=10.8$.
We combine these measurements with the low-mass stacking analysis of \citet{Henry13b}, 
who measure the $\log(F_{\mathrm{[OIII]}}/F_{\mathrm{H\beta}})$ ratio
at $\log(\mathrm{M_{\star}/M_{\odot}})=8.2-9.8$.  As we do not reach the same stellar mass limits as \citet{Henry13b} and 
the line ratios may behave differently below
$\log(\mathrm{M_{\star}/M_{\odot}})<8.5$, we ignore their lowest mass bin here.
We find that $\log(F_{\mathrm{[OIII]}}/F_{\mathrm{H\beta}})$ has a weak mass dependence,
which we use to derive the following correlation between the average ratio
of the [OIII] to H$\alpha$+[NII] $W$ and stellar mass:

\begin{equation}
\frac{W_\mathrm{{[OIII]}}}{W_\mathrm{{H\alpha+[NII]}}} = 0.95-0.46\left(\frac{\log\mathrm{M_{\star}/M_{\odot}}}{10^{10}}\right).
\label{eq:EWOIII}
\end{equation}

To convert the observed $\log(F_{\mathrm{[OIII]}}/F_{\mathrm{H\beta}})$ ratios
from H$\beta$ to H$\alpha$, we have assumed the most conservative case of a dust correction for case B recombination 
where $F_\mathrm{{H\alpha}} = 2.86 F_{\mathrm{H\beta}}$.  
When extrapolating the $W$ ratio from the emission line flux ratio, we must additionally make assumptions about the
average continuum flux near [OIII] and H$\alpha$ as 
$W_\mathrm{{[OIII]}}/W_\mathrm{{H\alpha}} = F_{\lambda\mathrm{[OIII]}}/F_{\lambda\mathrm{H\alpha}} \times F^{c}_{\lambda\mathrm{H\alpha}}/F^{c}_{\lambda\mathrm{[OIII]}}$.
We select galaxies at $z\sim1.4$ where H$\alpha$ falls cleanly into the $H_{F160W}$ broadband filter and [OIII] in the $J_{F125W}$ broadband filter,
finding an average $J_{F125W}-H_{F160W}$ color of -0.36 ABmag.
Although the argument quickly becomes circular with emission-line contamination in the broadband
filters, the contribution is expected to be small enough due to the broad filter width such that the average ratio can be used as a proxy for
${F^{c}_{\lambda\mathrm{H\alpha}}}/{F^{c}_{\lambda\mathrm{[OIII]}}}$ and folded into the constants in 
Equation~\ref{eq:EWOIII}.  The $W_{\mathrm{[OIII]}}$
is equal to the $W_{\mathrm{H\alpha+[NII]}}$ at $\log(\mathrm{M_{\star}/M_{\odot}})\sim10$, with lower values for more
massive galaxies and higher values for less massive galaxies.
As H$\beta$ falls in the same broadband filter as [OIII], we adopt the same correction for the continuum flux ratio to convert
from the $W_\mathrm{H\alpha+[NII]}$ to $W_\mathrm{H\beta}$.

\citet{Henry13b} find that the O32 ratio (${F_{\mathrm{[OIII]}}/F_{\mathrm{[OII]}}}$) does not have a strong 
mass dependence at $\log(\mathrm{M_{\star}/M_{\odot}})=9-10$.
When combining their measured value of O32 with the average $Z_{F814W}-J_{F125W}$ color of -0.65 mag, we assume that
$W_\mathrm{{[OII]}}/W_{\mathrm{[OIII]}}\sim0.5$.

The average flux density measured through a bandpass can be approximated as
$F_{\lambda} \simeq F^{c}_{\lambda}+F_{\mathrm{line}}/\Delta\lambda$, if we
assume a roughly flat continuum \citep{Papovich01}.
We calculate the width of the bandpass $\Delta\lambda$ by integrating the filter transmission curves.  The continuum
flux with emission-line contamination removed therefore becomes,

\begin{equation}
{F^{c}_{\lambda}} \simeq F_{\lambda}\left(\frac{1}{1+\frac{W_{\mathrm{rest}}(1+z)}{\Delta\lambda}}\right),
\label{eq:contam}
\end{equation}

\noindent where the best-fit mass-dependent relations for rest-frame $W$ of H$\alpha$+[NII], [OIII], [OII] and H$\beta$ are introduced above.
We can now correct the observed photometry for the entire sample for the average predicted emission-line contamination.
Where the photometric filters overlap with one of the four emission lines, the catalog fluxes will therefore become fainter by
$\Delta m \simeq -2.5\log(1+W_{\mathrm{rest}}(1+z)/\Delta\lambda)$.  We then re-run the stellar population synthesis
modeling with the same parameter settings as detailed in \citet{Skelton14} to derive new emission-line contamination
corrected stellar masses for our sample of star-forming galaxies.  The offsets applied to the photometry for the
four emission lines is a relatively strong
function of both stellar mass and redshift, where the largest corrections are applied to low-mass galaxies at the highest
redshifts.  For example, the average correction at $2.0<z<2.5$ for $\log(\mathrm{M_{\star}/M_{\odot}})=9$ is [OIII]$\sim0.7$ mag and H$\alpha\sim0.4$ mag.

We compare the change in stellar mass when estimating the increasing contribution to the broadband flux from emission lines
in Figure~\ref{fig:comparemass}.  We find that the stellar masses agree within $\lesssim0.05$ dex at all redshifts
for $\log(\mathrm{M})>10.0$.  The transparent points in Figure~\ref{fig:comparemass} indicate data with mass incompleteness.
The amount of contamination will depend on the number of bands included in the fit, their respective width,
and the redshift of the object: for surveys with a large number of bands the effect of emission
line contamination will be somewhat washed out; however, for surveys with fewer bands this effect can be
quite significant \citep[e.g.,][]{Kriek13}. The COSMOS and GOODS-S photometric catalogs include a large number of optical medium-band
filters, and COSMOS and partial coverage of AEGIS include NIR medium-band filters.
It follows from Equation~\ref{eq:contam} that the contamination will be most significant in cases where the emission lines
fall within these medium-band filters.
Although we find that the average contamination to the stellar masses agrees between the five fields, and hence is not sensitive
to the number of bands,
we note that all photometric catalogs have a significant number of photometric bands included.  The number of filters included ranges
from 18 broadband filters in UDS upwards to 44 broadband and medium-band filters in the COSMOS field.  Studies which incorporate far
fewer filters may suffer more severely from emission-line contamination.

In reality, the true contamination depends most sensitively on $W(z,\mathrm{M_{\star}})$.
As we are only interested in the emission-line contamination of the stellar masses on average, our assumptions for
the mass-dependencies of $W$ are suitable.  However, we caution the reader when applying the average relation presented in Figure~\ref{fig:comparemass}
and Table~\ref{tab:contam} to individual galaxies that deviate significantly from the average galaxy at a given redshift and stellar mass.
For example, a significantly larger mass contamination may occur for high redshift galaxies with the highest sSFRS \citep[e.g.,][]{Atek11}.

As there is negligible contamination at high stellar masses, there is no significant change to the best-fit relation 
presented in Section~\ref{sec:powerlaw}.
The formal best-fit for the redshift evolution of the low mass slope changes from $\alpha(z) = 0.87\pm0.06 + (0.11\pm0.04)z$ 
to $\alpha(z) = 0.95\pm0.05 -  (0.02\pm0.04)z$, before and
after accounting for contamination to the stellar masses from emission lines.
The predicted emission line contamination to the stellar
masses shifts the low-mass slopes shallower by 0.01--0.16, changing the average value from $\alpha=1.03\pm0.08$ to $\alpha=0.97\pm0.06$.
Although the contamination starts to become significant below $\log(\mathrm{M_{\star}/M_{\odot}})\sim9$,
the measured $\Delta\log\mathrm{M_{star}}$ values in Table~\ref{tab:contam} would need to be underestimated by a
factor of 5 at $1.5<z<2.0$ and a factor of 2 at $2.0<z<2.5$ to
remove the observed trends of a steeper slope at low masses. 
Figure~\ref{fig:comparemass} shows only a weak mass dependence for the
emission line contamination at $z<1.5$.  The only way to overestimate the low-mass slope at these redshifts would therefore be to assume
similar equivalent widths to those observed at $z>1.5$.  We therefore conclude that although contamination
from emission lines result in a significant correction to the stellar mass, slopes measured without the 
emission line contamination removed are robust.

\begin{table*}[ht!]
\centering
\begin{threeparttable}
    \caption{Estimated Emission-line Contamination to Stellar Mass}\label{tab:mass}
    \begin{tabular}{lcccc}
      \hline \hline
      \noalign{\smallskip}
$\log(\mathrm{M_{\star}/M_{\odot}})$~~~~~~     & $0.5<z<1.0$ & $1.0<z<1.5$ & $1.5<z<2.0$ & $2.0<z<2.5$ \\
      \noalign{\smallskip}
      \hline
      \noalign{\smallskip}
      8.0--8.2 & $ 0.090\pm 0.005$ & $ 0.232\pm 0.008$ & -- & --  \\
      8.2--8.4 & $ 0.077\pm 0.004$ & $ 0.170\pm 0.006$ & $ 0.337\pm 0.007$ & $ 0.504\pm 0.022$ \\
      8.4--8.6 & $ 0.062\pm 0.004$ & $ 0.119\pm 0.004$ & $ 0.295\pm 0.005$ & $ 0.507\pm 0.011$ \\
      8.6--8.8 & $ 0.049\pm 0.003$ & $ 0.084\pm 0.004$ & $ 0.231\pm 0.004$ & $ 0.483\pm 0.007$ \\
      8.8--9.0 & $ 0.035\pm 0.003$ & $ 0.050\pm 0.003$ & $ 0.192\pm 0.004$ & $ 0.439\pm 0.006$ \\
      9.0--9.2 & $ 0.031\pm 0.003$ & $ 0.022\pm 0.002$ & $ 0.152\pm 0.004$ & $ 0.361\pm 0.006$ \\
      9.2--9.4 & $ 0.026\pm 0.003$ & $ 0.006\pm 0.002$ & $ 0.094\pm 0.003$ & $ 0.283\pm 0.006$ \\
      9.4--9.6 & $ 0.030\pm 0.003$ & $ 0.004\pm 0.002$ & $ 0.059\pm 0.003$ & $ 0.187\pm 0.005$ \\
      9.6--9.8 & $ 0.027\pm 0.003$ & $ 0.004\pm 0.002$ & $ 0.033\pm 0.003$ & $ 0.123\pm 0.004$ \\
      9.8--10.0 & $ 0.026\pm 0.003$ & $ 0.004\pm 0.002$ & $ 0.016\pm 0.003$ & $ 0.088\pm 0.003$ \\
      10.0--10.2 & $ 0.017\pm 0.002$ & $ 0.007\pm 0.002$ & $ 0.011\pm 0.003$ & $ 0.070\pm 0.004$ \\
      10.2--10.4 & $ 0.010\pm 0.002$ & $ 0.009\pm 0.002$ & $ 0.008\pm 0.003$ & $ 0.049\pm 0.004$ \\
      10.4--10.6 & $ 0.011\pm 0.001$ & $ 0.013\pm 0.002$ & $ 0.001\pm 0.003$ & $ 0.037\pm 0.003$ \\
      10.6--10.8 & $ 0.006\pm 0.001$ & $ 0.010\pm 0.001$ & $ 0.004\pm 0.003$ & $ 0.031\pm 0.004$ \\
      10.8--11.0 & $ 0.004\pm 0.001$ & $ 0.013\pm 0.002$ & $ 0.006\pm 0.002$ & $ 0.023\pm 0.004$ \\
      11.0--11.2 & $ 0.003\pm 0.005$ & $ 0.004\pm 0.002$ & $ 0.006\pm 0.002$ & $ 0.025\pm 0.005$ \\
      11.2--11.4 & -- & -- & $ 0.005\pm 0.003$ & $ 0.033\pm 0.005$ \\
      \noalign{\smallskip}
      \hline
      \noalign{\smallskip}
    \end{tabular}
    \begin{tablenotes}
      \small
    \item \emph{Notes.} $\Delta\log\mathrm{M_{\star}}=\log\mathrm{M_{\star,before}}-\log\mathrm{M_{\star,after}}$ in
      logarithmic units of dex. Here, ``before'' signifies the original stellar masses and ``after'' signifies the new stellar masses when
        subtracting the estimated redshift and mass-dependent contaminating emission-line flux for [OII], H$\beta$, [OIII] and H$\alpha$
        in the relevant filters.
    \end{tablenotes}
    \label{tab:contam}
  \end{threeparttable}
\end{table*}

\section{Cosmic Variance of the Star Formation Sequence}

\begin{figure}[t]
\leavevmode
\centering
\includegraphics[width=\linewidth]{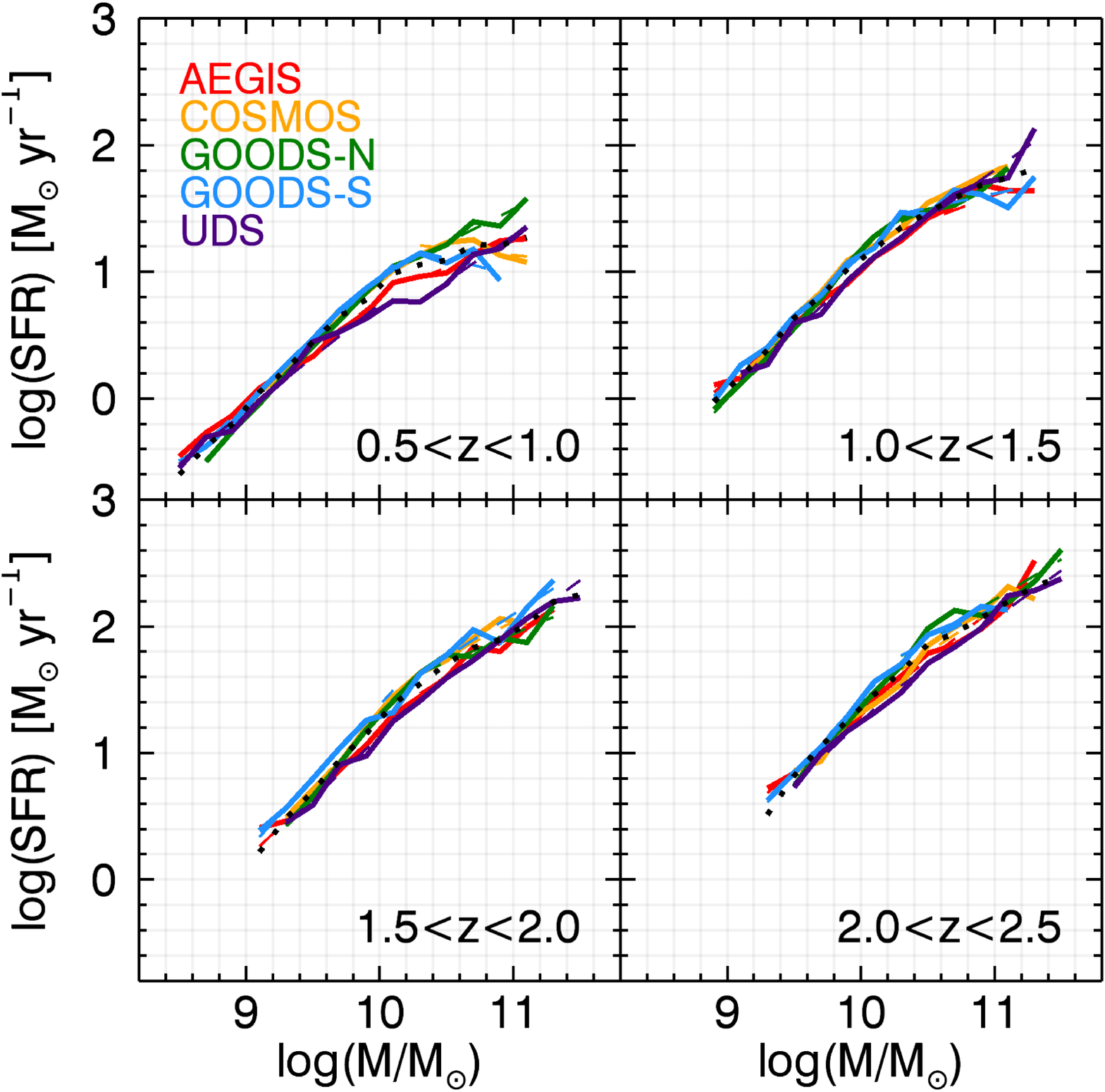}
\caption{The star formation rate as a function of stellar mass for star-forming galaxies, with UV+IR SFRs measured in
stacking analyses in the five individual CANDELS fields.}
\label{fig:sfr_fields}
\end{figure}

Measurements of the observable universe are affected by cosmic large-scale structure, where a measurement of any region of the sky 
may differ from that in a different region of the sky by an amount that may be much greater than the sample variance.
To combat the effects of this cosmic variance, it is preferable to combine all of the CANDELS fields together in the analysis
of the star formation sequence.  However, it is also informative to measure these relations
by repeating the stacking analysis detailed in Section~\ref{sec:sfr} for each of the five fields separately.  The median UV+IR SFRs
in 0.2 dex bins of stellar mass are shown in Figure~\ref{fig:sfr_fields}, color-coded by field. 
Given the steep turn-over of the stellar mass function at the highest masses \citep[e.g.,][]{Marchesini09}, 
the larger variations in the measured relation between the fields at the highest masses are not surprising.  The variations 
between the measured star formation sequence for the individual fields is remarkably small at all redshifts below $\log(\mathrm{M_{\star}/M_{\odot}})<10$.

\begin{figure}[t]
\leavevmode
\centering
\includegraphics[width=\linewidth]{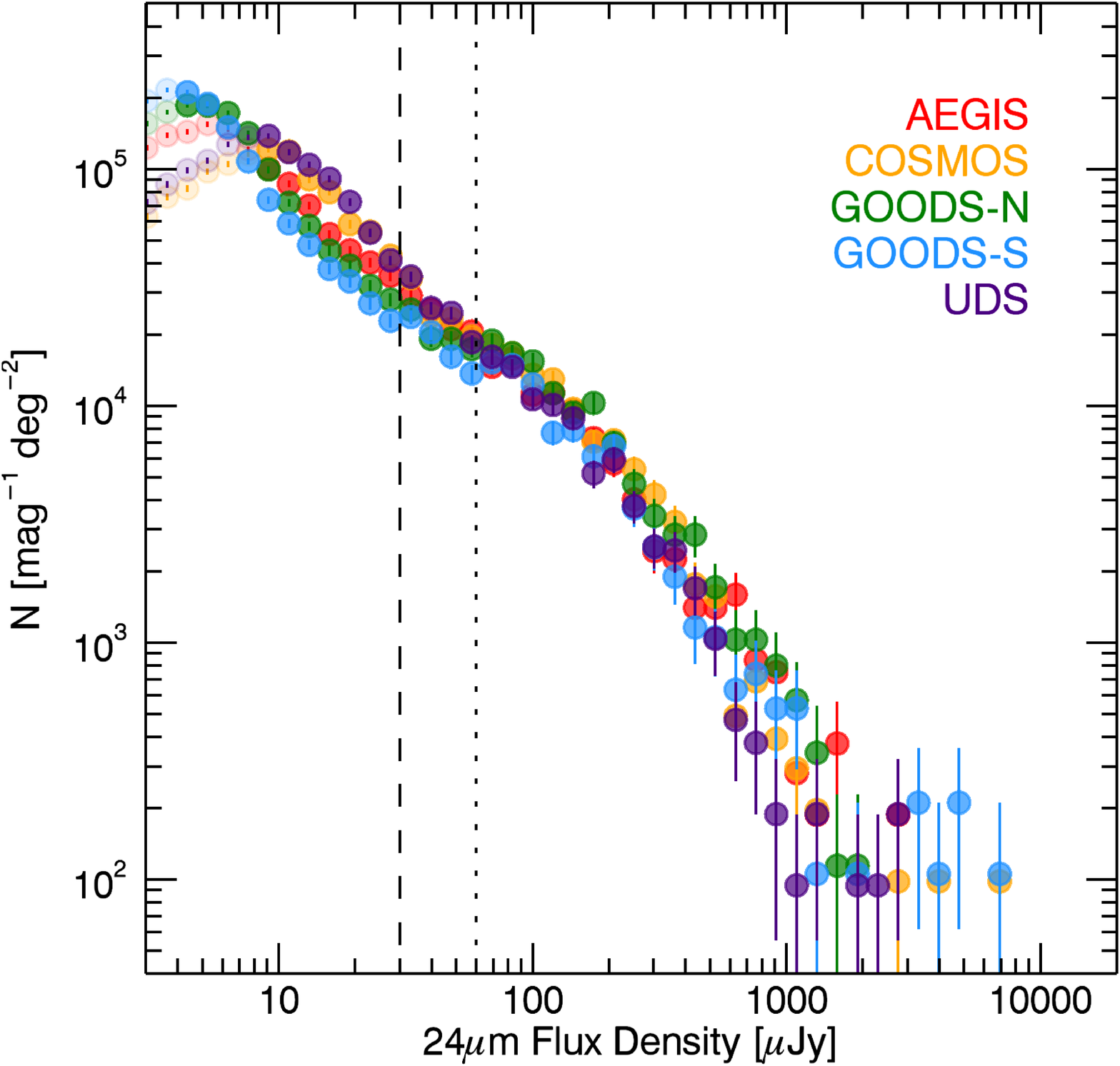}
\caption{Number of Spizter/MIPS 24$\mu$m detected objects per magnitude per square degree as a function of the total
24$\mu$m flux density in the five CANDELS fields.  The dashed line corresponds to the $3\sigma$ 24$\mu$m detection 
limit in the deeper GOODS fields, whereas the dotted line is the $3\sigma$ detection limit in the remaining three fields.
Measurements below the respective $1\sigma$ limits are shown with lighter color symbols.}
\label{fig:numcounts}
\end{figure}

Although we cannot rule out differences due solely to cosmic variance, the star formation sequence from the
UDS field does appear to lie below the other 
fields at all redshifts, hinting at calibration uncertainties in the data.  We perform zero point corrections 
to the optical to NIR photometric catalogs, as described in \citet{Skelton14}, but similar such tests are not 
possible with the Spitzer/MIPS 24$\mu$m photometry.  Figure~\ref{fig:numcounts} shows the number counts 
of Spitzer/MIPS 24$\mu$m objects in the five CANDELS fields.  Indeed, the 24$\mu$m number counts in 
the UDS field are generally higher at the faintest magnitudes and lower at the brightest magnitudes, 
relative to the other four fields.  Deviations larger than the sample variance, as reflected through the
Poisson error bars in Figure~\ref{fig:numcounts},
are generally attributed to cosmic variance.
These differences in the large-scale structure between the five fields may explain
the variations in the measured star formation sequence in Figure~\ref{fig:sfr_fields}, although we cannot rule
out additional calibration uncertainties.

In general, the 3D-HST Spitzer/MIPS 24$\mu$m photometry of detected sources 
in the five independent CANDELS fields agree well with measurements presented in other public catalogs.
We describe several comparisons to external photometric catalogs in detail in the following paragraphs.

We find that no systematic offsets exist between the 24$\mu$m photometry of the detected sources in 
the 3D-HST and the NMBS surveys in the COSMOS and AEGIS fields.  However, we note that although the 
higher-resolution prior is different, both surveys rely on the same base Spitzer/MIPS data products and analysis codes.  

The public SpUDS MIPS 24$\mu$m photometric catalog contains aperture photometry within an aperture of radius 7.5$^{\prime\prime}$,
with an aperture correction to total flux density and a minimum flux density cut of 300$\mu$Jy.   Only 88 bright sources
above these limits overlap with the UDS/CANDELS field, and the flux density measurements are on average 0.15 mag fainter
in the 3D-HST catalogs.  This offset may be attributed to the different aperture photometry techniques and
aperture corrections adopted.  As no further details are published regarding the SpUDS photometry, we
cannot further investigate the cause of these discrepancies.

The 3D-HST 24$\mu$m photometry in GOODS-N is well matched to the public MODS catalogs \citep{Kajisawa11}, with 
no systematic offsets and agreement between the two catalogs on order $<0.05$ mag above 30$\mu$Jy.  
We also compare our photometry in both the GOODS-S and GOODS-N fields to the photometric catalogs of \citet{Teplitz11}.  
The 3D-HST MIPS 24$\mu$m flux densities are systematically fainter by 0.06 magnitudes in both fields, 
after removing a small color correction they have applied to account for instrument calibrations. 

\section{Quantifying the Star Formation Sequence for All Galaxies}

\begin{figure*}[t]
\leavevmode
\centering
\plottwo{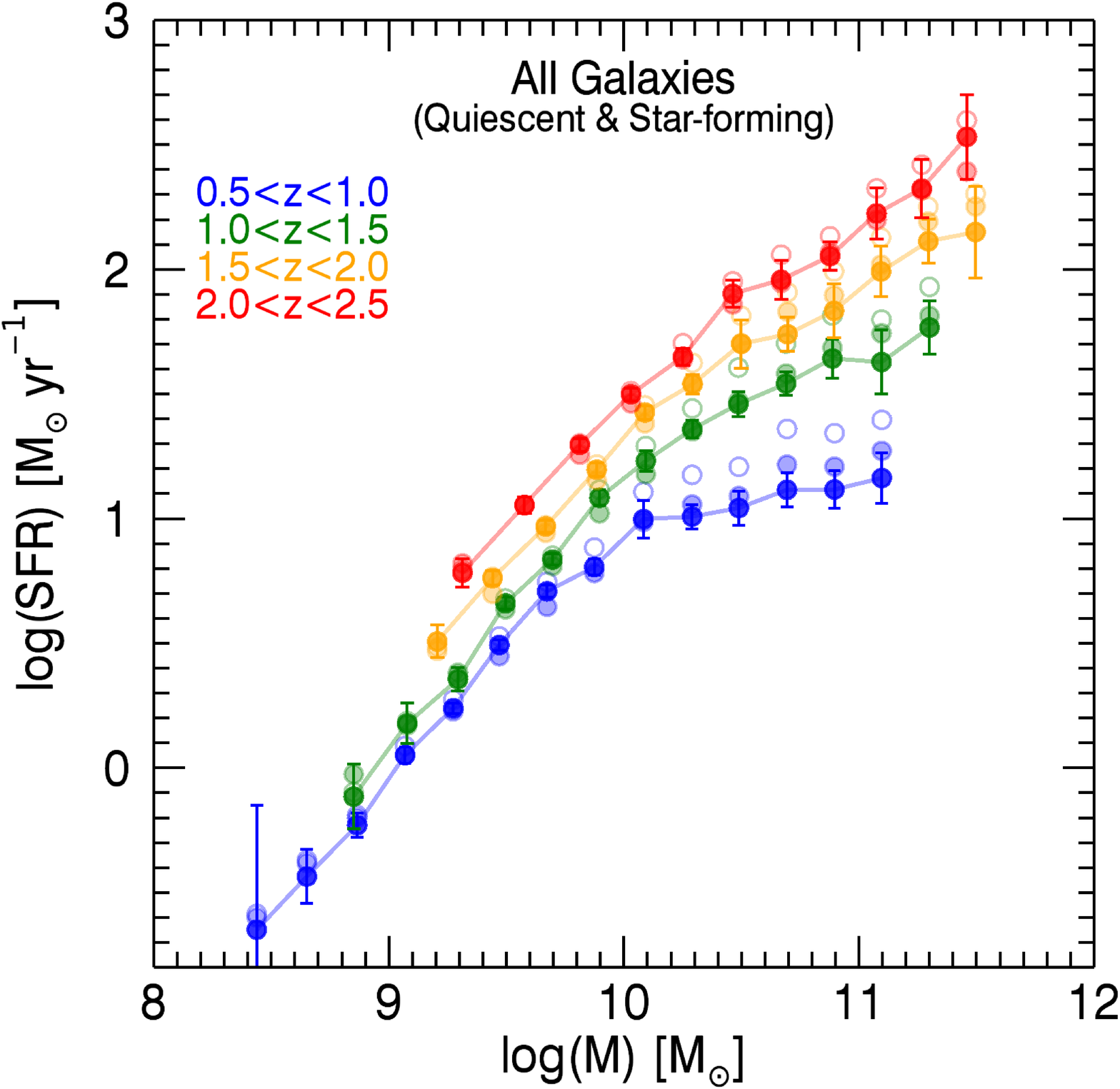}{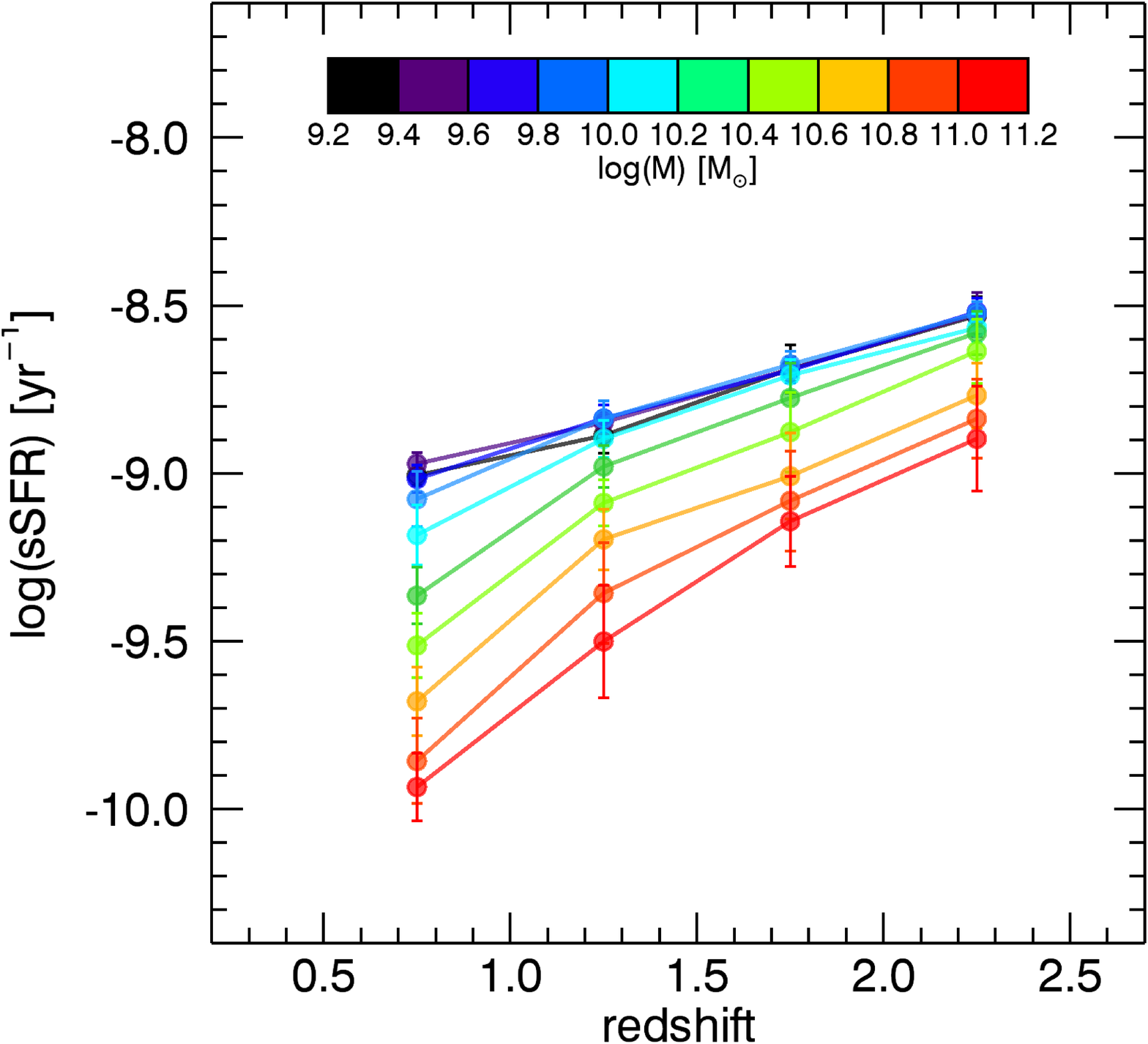}
\caption{The observed $\log\mathrm{M_{\star}}$ as a function of $\log\Psi$ (left), and
the mass-dependent redshift evolution of the sSFR (right) for all galaxies, both star-forming and quiescent.  
Average stacks of all galaxies are slightly lower at the highest masses, but
generally similar to the median stacking analysis of star-forming galaxies only.}
\label{fig:sams}
\end{figure*}

We have repeated the Spitzer/MIPS 24$\mu$m stacking analysis detailed in Section~\ref{sec:sfr} for all galaxies.
Figure~\ref{fig:sams} shows the $\log\Psi-\log\mathrm{M_{\star}}$ and $\log(\Psi/\mathrm{M_{\star}})-z$ relations.
The only difference here is that we add the quiescent galaxies that were removed through 
our rest-frame color selection described in Section~\ref{sec:selection} back into the sample.
We provide the stacking analysis results for all galaxies in Table~\ref{tab:sfr_all}, 
including the average measured SFRs, $\mathrm{L_{IR}}$, and $\mathrm{L_{UV}}$ for all redshift and stellar mass bins.

\begin{table*}[ht]
\centering
\begin{threeparttable}
    \caption{Star Formation Sequence Data: All Galaxies}\label{tab:sfr_all}
    \begin{tabular}{lcccc}
      \hline \hline
      \noalign{\smallskip}
      & ~~~$\log\mathrm{M}_{\star}$~~~ & ~~~~~$\log\Psi$~~~~~ & ~~~~~$\log\mathrm{L_{IR}}$~~~~~ & ~~~~~$\log\mathrm{L_{UV}}$~~~~~ \\
      \noalign{\smallskip}
      \hline
      \noalign{\smallskip}
      $0.5<z<1.0$~~~  &  8.7 &  $-0.44\pm 0.11$ & $  8.40\pm  0.43$ & $  9.15\pm  0.01$ \\
             &  8.9 &  $-0.23\pm 0.05$ & $  8.97\pm  0.07$ & $  9.31\pm  0.01$ \\
             &  9.1 &  $ 0.05\pm 0.03$ & $  9.61\pm  0.03$ & $  9.46\pm  0.01$ \\
             &  9.3 &  $ 0.24\pm 0.02$ & $  9.88\pm  0.01$ & $  9.58\pm  0.01$ \\
             &  9.5 &  $ 0.49\pm 0.02$ & $ 10.25\pm  0.01$ & $  9.69\pm  0.01$ \\
             &  9.7 &  $ 0.71\pm 0.02$ & $ 10.53\pm  0.01$ & $  9.77\pm  0.01$ \\
             &  9.9 &  $ 0.81\pm 0.03$ & $ 10.67\pm  0.02$ & $  9.72\pm  0.02$ \\
             & 10.1 &  $ 1.00\pm 0.07$ & $ 10.91\pm  0.02$ & $  9.69\pm  0.08$ \\
             & 10.3 &  $ 1.01\pm 0.05$ & $ 10.92\pm  0.02$ & $  9.64\pm  0.02$ \\
             & 10.5 &  $ 1.04\pm 0.07$ & $ 10.97\pm  0.02$ & $  9.54\pm  0.05$ \\
             & 10.7 &  $ 1.12\pm 0.07$ & $ 11.04\pm  0.02$ & $  9.67\pm  0.02$ \\
             & 10.9 &  $ 1.12\pm 0.08$ & $ 11.04\pm  0.03$ & $  9.64\pm  0.03$ \\
             & 11.1 &  $ 1.16\pm 0.10$ & $ 11.08\pm  0.04$ & $  9.77\pm  0.04$ \\
      \noalign{\smallskip}
      \hline
      \noalign{\smallskip}
      $1.0<z<1.5$~~~  &   9.1 &  $ 0.18\pm 0.08$ & $  9.58\pm  0.10$ & $  9.66\pm  0.01$ \\
             &  9.3 &  $ 0.36\pm 0.05$ & $  9.86\pm  0.06$ & $  9.79\pm  0.01$ \\
             &  9.5 &  $ 0.66\pm 0.02$ & $ 10.39\pm  0.02$ & $  9.90\pm  0.01$ \\
             &  9.7 &  $ 0.84\pm 0.02$ & $ 10.63\pm  0.02$ & $  9.96\pm  0.01$ \\
             &  9.9 &  $ 1.08\pm 0.03$ & $ 10.95\pm  0.02$ & $  9.99\pm  0.03$ \\
             & 10.1 &  $ 1.23\pm 0.04$ & $ 11.13\pm  0.02$ & $  9.99\pm  0.04$ \\
             & 10.3 &  $ 1.36\pm 0.04$ & $ 11.28\pm  0.02$ & $  9.88\pm  0.03$ \\
             & 10.5 &  $ 1.46\pm 0.05$ & $ 11.40\pm  0.02$ & $  9.83\pm  0.04$ \\
             & 10.7 &  $ 1.54\pm 0.05$ & $ 11.48\pm  0.02$ & $  9.85\pm  0.02$ \\
             & 10.9 &  $ 1.64\pm 0.08$ & $ 11.59\pm  0.04$ & $  9.90\pm  0.02$ \\
             & 11.1 &  $ 1.63\pm 0.13$ & $ 11.57\pm  0.07$ & $  9.99\pm  0.04$ \\
             & 11.3 &  $ 1.77\pm 0.11$ & $ 11.71\pm  0.08$ & $ 10.09\pm  0.03$ \\
      \noalign{\smallskip}
      \hline
      \noalign{\smallskip}
      $1.5<z<2.0$~~~  &  9.2 &  $ 0.51\pm 0.07$ & $ 10.01\pm  0.07$ & $  9.94\pm  0.01$ \\
             &  9.4 &  $ 0.76\pm 0.03$ & $ 10.45\pm  0.02$ & $ 10.05\pm  0.01$ \\
             &  9.7 &  $ 0.97\pm 0.02$ & $ 10.73\pm  0.02$ & $ 10.16\pm  0.01$ \\
             &  9.9 &  $ 1.20\pm 0.03$ & $ 11.05\pm  0.02$ & $ 10.18\pm  0.02$ \\
             & 10.1 &  $ 1.43\pm 0.03$ & $ 11.33\pm  0.01$ & $ 10.15\pm  0.02$ \\
             & 10.3 &  $ 1.54\pm 0.04$ & $ 11.46\pm  0.02$ & $ 10.09\pm  0.02$ \\
             & 10.5 &  $ 1.70\pm 0.10$ & $ 11.64\pm  0.02$ & $  9.95\pm  0.16$ \\
             & 10.7 &  $ 1.74\pm 0.07$ & $ 11.69\pm  0.03$ & $  9.92\pm  0.04$ \\
             & 10.9 &  $ 1.83\pm 0.11$ & $ 11.78\pm  0.04$ & $  9.95\pm  0.12$ \\
             & 11.1 &  $ 1.99\pm 0.10$ & $ 11.94\pm  0.05$ & $ 10.04\pm  0.03$ \\
             & 11.3 &  $ 2.11\pm 0.09$ & $ 12.06\pm  0.05$ & $ 10.20\pm  0.04$ \\
             & 11.5 &  $ 2.15\pm 0.18$ & $ 12.10\pm  0.18$ & $ 10.26\pm  0.06$ \\
      \noalign{\smallskip}
      \hline
      \noalign{\smallskip}
      $2.0<z<2.5$~~~  &  9.3 &  $ 0.78\pm 0.06$ & $ 10.33\pm  0.08$ & $ 10.19\pm  0.01$ \\
             &  9.6 &  $ 1.05\pm 0.03$ & $ 10.76\pm  0.03$ & $ 10.32\pm  0.02$ \\
             &  9.8 &  $ 1.30\pm 0.03$ & $ 11.10\pm  0.02$ & $ 10.41\pm  0.01$ \\
             & 10.0 &  $ 1.50\pm 0.03$ & $ 11.36\pm  0.01$ & $ 10.42\pm  0.02$ \\
             & 10.3 &  $ 1.65\pm 0.03$ & $ 11.56\pm  0.02$ & $ 10.33\pm  0.03$ \\
             & 10.5 &  $ 1.90\pm 0.05$ & $ 11.84\pm  0.02$ & $ 10.30\pm  0.05$ \\
             & 10.7 &  $ 1.96\pm 0.08$ & $ 11.90\pm  0.03$ & $ 10.13\pm  0.07$ \\
             & 10.9 &  $ 2.05\pm 0.06$ & $ 12.00\pm  0.04$ & $ 10.14\pm  0.02$ \\
             & 11.1 &  $ 2.22\pm 0.10$ & $ 12.18\pm  0.04$ & $ 10.18\pm  0.08$ \\
             & 11.3 &  $ 2.32\pm 0.12$ & $ 12.28\pm  0.07$ & $ 10.18\pm  0.06$ \\
             & 11.5 &  $ 2.53\pm 0.17$ & $ 12.49\pm  0.09$ & $ 10.22\pm  0.11$ \\
      \noalign{\smallskip}
      \hline
      \noalign{\smallskip}
    \end{tabular}
    \begin{tablenotes}
      \small
      \item \emph{Notes.} All galaxies (both quiescent and star-forming) are included in the stacking analyses.
        Stellar masses are in units of M$_{\odot}$ and include a correction for emission-line contamination, as detailed in Appendix A.
        Star formation rates are in units of M$_{\odot}$ yr$^{-1}$.
        Luminosities are in units of L$_{\odot}$. $\mathrm{L_{IR}}$ is the bolometric FIR luminosity as calibrated
        from the average 24$\mu$m stacks, and $\mathrm{L_{UV}}$ (or $1.5\nu\mathrm{L_{\nu,2800}}$) is the average rest-frame UV luminosity at
        1216--3000$\mathrm{\AA}$.
    \end{tablenotes}
  \end{threeparttable}
\end{table*}

\addcontentsline{toc}{chapter}{\numberline {}{\sc References}}

\end{document}